\documentclass[12pt]{article}
\usepackage{graphicx}
\usepackage{amssymb}
\usepackage{amsmath}
\usepackage{color,cite}
\usepackage{tikz,xcolor,dsfont,scalefnt}
\usetikzlibrary{calc,decorations.markings,intersections,shapes.arrows,decorations.markings,snakes}

\allowdisplaybreaks

\newcommand{\poincare}{Poincar\'e }
\newcommand{\cW}{ {\mathcal W} }
\newcommand{\cP}{ {\mathcal P} }
\newcommand{\nn}{\nonumber}
\newcommand{\als}{\alpha_s}
\newcommand{\ba}{\bar{\alpha}}

\definecolor{darkred}{rgb}{0.7,0.0,0.0}
\newcommand{\red}{\color{darkred}}
\newcommand{\Li}{ {\mathrm{Li}} }
\newcommand{\az}{ \Phi }
\newcommand{\cI}{ {\mathcal I} }

\newcommand{\cS}{ {\mathcal S}}
 \newcommand{\pslr}{ {\rm PSL}(2,\mathbb{R}) }
 \newcommand{\pslc}{ {\rm PSL}(2,\mathbb{C}) }
\newcommand{\tL}{ {L }}
\newcommand{\hL}{ {\widehat{L}} }
\newcommand{\met}[1]{{\red \langle}{#1}  {\red \rangle}}
\newcommand{\nbar}{ \overline{\hspace{-0.2mm} n\hspace{-0.2mm}}}

\begin{document} 

\catcode`\@=11
\font\manfnt=manfnt
\def\Watchout{\@ifnextchar [{\W@tchout}{\W@tchout[1]}}
\def\W@tchout[#1]{{\manfnt\@tempcnta#1\relax%
  \@whilenum\@tempcnta>\z@\do{%
    \char"7F\hskip 0.3em\advance\@tempcnta\m@ne}}}
\let\foo\W@tchout
\def\dubious{\@ifnextchar[{\@dubious}{\@dubious[1]}}
\let\enddubious\endlist
\def\@dubious[#1]{%
  \setbox\@tempboxa\hbox{\@W@tchout#1}
  \@tempdima\wd\@tempboxa
  \list{}{\leftmargin\@tempdima}\item[\hbox to 0pt{\hss\@W@tchout#1}]}
\def\@W@tchout#1{\W@tchout[#1]}
\catcode`\@=12

\thispagestyle{empty}

\hfill\begin{tabular}{@{}p{.5\linewidth}@{}}
\multicolumn{1}{@{}c@{}}{SLAC--PUB--15920} \\
\end{tabular}

 \vspace{5mm}

\begingroup\centering
{\Large\bfseries\mathversion{bold}
Non-global Logarithms\\[3mm] at 3 Loops, 
4 Loops, 5 Loops and Beyond
}%

\vspace{8mm}

\begingroup\scshape\large
Matthew~D.~Schwartz$^{(1)}$, Hua~Xing~Zhu$^{(2)}$\\
\endgroup
\vspace{5mm}
\begingroup\small
$^{(1)}$\emph{Center for the Fundamental Laws of Nature,
Harvard University}\\ 
$^{(2)}$\emph{SLAC National Accelerator Laboratory,
Stanford University, Stanford, CA 94309, USA} 
\endgroup

\vspace{0.4cm}
\begingroup\small
E-mails:
{\tt schwartz@physics.harvard.edu}, {\tt hxzhu@slac.stanford.edu}\endgroup
\vspace{0.7cm}

\vspace{10mm}
\textbf{Abstract}\vspace{5mm}\par
\begin{minipage}{14.7cm}
We calculate the coefficients of the leading non-global logarithms for the
hemisphere mass distribution analytically at $3$, $4$, and $5$ loops at large $N_c$.
We confirm that the integrand derived with the strong-energy-ordering approximation
and fixed-order iteration of the Banfi-Marchesini-Syme~(BMS) equation agree.
Our calculation exploits a hidden $\pslr$ symmetry associated with the jet directions,
apparent in the BMS equation after a stereographic projection to the \poincare disk.
The required integrals have an iterated form, leading to functions of uniform transcendentality. 
This allows us to extract the coefficients, and some functional dependence on the jet directions,
by computing the symbols and coproducts of appropriate expressions involving classical
and Goncharov polylogarithms.
Convergence of the series to a numerical solution of the BMS equation is also discussed.

\end{minipage}\par
\endgroup

\newpage

\section{Introduction}
\label{sec:intro}
Jet substructure is playing an increasingly prominent role in the physics of
hadron collisions, particularly at the LHC~\cite{0802.2470, 0806.0848, 
1010.3698,1011.2268,1012.2077,1106.3076,1201.1914,1201.0008 }.
As new substructure methods are developed, it will be important to cross check
the approximations used in Monte Carlo simulations both against data and against independent precision calculations in QCD.
Such substructure calculations necessarily involve resummation of large logarithms. For sufficiently inclusive
observables, this resummation is relatively straightforward. 
 For
example, traditional jet mass and shape distribution have been studied
in great detail, both at $e^+e^-$
colliders~\cite{0912.0262,1001.0014,1004.3483,1005.1644,1102.0561,1105.3676,1105.4628,1110.0004,1111.2016,1112.3343,1309.3560}
and hadron
colliders~\cite{0911.0681,1008.4355,1106.4310, 1107.4535, 1204.3898,1206.1344,1207.1640,1206.6115,1208.0010,1212.2106,1302.0846,1303.6637,1305.0007,1307.0007,1307.0013,1401.4458,1402.2657}.
However, for calculations involving multiple scales and
parameters, such as with extra electroweak gauge bosons, jet vetoes, or jet algorithm parameters, it is less clear how to guarantee
that all the relevant large logarithms are resummed. One particular challenge is to understand non-global 
logarithms (NGLs)~\cite{Dasgupta:2001sh,hep-ph/0203009,hep-ph/0208073,Banfi:2002hw,hep-ph/0211426,hep-ph/0308284,1002.4557}.

Non-global logarithms were
first characterized and understood by Dasgupta and Salam (DS) ~\cite{Dasgupta:2001sh}. They arise in
exclusive observables for which phase space cuts unequally distribute the real
and virtual contributions. Consider for example, the doubly differential
distribution in hemisphere masses, $\frac{d^2 \sigma}{d m_L d m_R}$ in an $e^+
e^-$ collision~\cite{1105.3676,1105.4628}. In certain limits, such as when  $m_L \gg
m_R$~(or vice versa), NGLs of the form $L = \left|
  \ln\frac{m_L}{m_R}\right|$ can give large contributions to the distribution. 
The leading dependence of the hemisphere mass cross section on $L$ (of order  $\alpha_s^2 L^2$), was
 computed in~\cite{Dasgupta:2001sh}. Subleading NGLs (of order $\alpha_s^2 L$)  were first computed
in~\cite{1105.3676,1105.4628}; these calculations also revealed surprising~\cite{0806.3852} non-logarithmic dependence on the non-global ratio $\frac{m_L}{m_R}$.

Whether NGLs provide a quantitatively important contribution to precision calculations, and whether they must be resummed, is only beginning to 
be understood~\cite{1102.0561,1111.2016,1207.1640,1208.0010,1302.0846}.
Although global logarithms in substructure observables can be resummed using the renormalization group, for example using
Soft-Collinear Effective Theory~\cite{hep-ph/0011336,hep-ph/0109045,hep-ph/0206152}, it is not known how (or if) NGLs can be resummed
using similar methods. Fortunately, the leading NGLs (terms of the form $(\alpha L)^n$) can be reproduced within the strong-energy-ordering approximation to QCD.
This approximation leads to simplified cross sections, particularly at large $N_c$, and allows for straightforward
resummation using Monte-Carlo~(MC) simulation~\cite{Dasgupta:2001sh}.
As an alternative to the MC approach, Banfi, Marchesini and Syme (BMS) have derived, using the same approximations, 
integro-differential equation~\cite{Banfi:2002hw}, which describes the
evolution of leading NGLs at large $N_c$. Remarkably, the BMS equation is mathematically
similar to the Balitsky-Kovchegov~(BK) equation describing the
dynamics of gluon saturation at small
$x$~\cite{NUPHA.B463.99,PHRVA.D60.034008}. Based on this formal
similarity, a finite $N_c$ generalization of BMS equation has been
proposed in Ref.~\cite{hep-ph/0312050}, and numerically studied in Ref.~\cite{1304.6930}.

\enlargethispage{20pt}

The BMS equation has the potential not just to resum the NGLs, but also to give us insights into their structure and importance.
An interesting feature of this equation is that it has a hidden $\pslr$ symmetry. More precisely, let us define the hemispheres
with respect to the $n$ direction. Then consider the contribution $g_{ab}(L)$ to the right-hemisphere-mass distribution from a dipole,
that is a pair of rapidly moving colored particles, in the $a$ and $b$ directions (neither of which are necessarily aligned with the left-hemisphere axis $\bar{n}^\mu$). While
one would naturally expect cylindrical symmetry as $a$ and $b$ rotate around the hemisphere axis, there is actually a much larger $\pslr$ symmetry
acting on $a$ and $b$. 

To calculate the $n$-loop leading NGL, we can iterate the BMS equation to produce the correct integrand. This iteration
is equivalent to, but significantly simpler then, summing the relevant real, virtual, and real-virtual contributions 
in the strongly ordered limit and then subtracting the global contribution.
Exploiting the $\pslr$ symmetry, the calculation of $g_{ab}(L)$ for arbitrary $a$ and $b$ simplifies. Furthermore, since
the integrands have an iterated structure, we can exploit the technology of symbols and coproducts to simplify expressions involving
polylogarithms. Our final result for the leading NGL at 5 loops is
\begin{equation}
  g_{n\nbar}(\hL) = -\frac{\pi^2}{24} \hL ^2 + \frac{\zeta(3)}{12} \hL^3 +
  \frac{\pi^4}{34560} \hL^4 + \left( -\frac{\pi^2 \zeta(3)}{360} +
  \frac{17\zeta(5)}{480} \right)\hL^5 + \dots
\end{equation}
which is to be evaluated at $\hL \equiv N_c \frac{\als}{\pi} \ln\frac{m_L}{m_R}$.

The paper is organized as follows. Section ~\ref{sec:global} gives a review of how to define the non-global contribution precisely, particularly
in the hemisphere case.  Section~\ref{sec:seo} reviews the strong energy ordering approximation and the simplifications in engenders at large $N_c$. 
Section~\ref{sec:nghmi} shows how the integrand for the hemisphere mass distribution can be derived using strong-energy-ordering, including both real and
virtual contributions. Although the procedure is systematic, it becomes quite involved already at 3 loops.  Section~\ref{sec:BMS} presents the BMS
 equation for the NGLs in the hemisphere mass distribution. We do not rederive BMS. Instead we check that when expanded to fixed order it gives
exactly the SEO integrand including both real and virtual contributions. Section~\ref{sec:simp} simplifies the BMS equation. In particular, in this
section we show that it respects the $\pslr$ symmetry of the \poincare disk which drastically simplifies the perturbative calculation. Section~\ref{sec:ngl2}
begins our perturbative calculation of the NGLs to 3, 4 and 5 loops. The methods we employ include contour integration as well as the use of Goncharov polylogarithms,
symbols and coproducts. 
Section~\ref{sec:resum} discusses how to solve the BMS equation numerically, to all orders in $\als$. We compare our solution to that of
Dasgupta and Salam, finding very good agreement. We also compare the resummed distribution to the perturbative series and to various approximations.
Section~\ref{sec:finitenc} discusses possible generalizations to finite $N_c$ and we conclude in Section~\ref{sec:conclude}.

\section{Global and non-global logs}
\label{sec:global}
Two facts make the resummation of the leading non-global logarithm tractable.
First, these logarithms can only come from regions of real or virtual phase
space where the gluons are strongly ordered in energy. Second,
cross-sections in QCD simplify in the strong-energy ordered limit, particularly at large $N_c$.

In this paper, we are mainly interested in the hemisphere mass distribution in
$e^+ e^- \rightarrow$ jets. A precise calculation of this observable is
relevant to precision physics both at $e^+ e^-$ colliders and indirectly at
hadron colliders~\cite{1207.1640,1208.0010}. We work in the dijet limit, where the jets are back-to-back in the
$n^{\mu} = ( 1, \vec{n} )$ and $\nbar^{\mu} = ( 1, - \vec{n})$ directions. These directions define the hemisphere axis. Our convention is that $\vec{n}$ defines the right-hemisphere axis. Let $m_L$
and $m_R$ be the left and right hemisphere masses respectively and $Q$ be the
center-of-mass energy. In the dijet limit, we have $m_L \ll Q$ and
$m_R \ll Q$.

As is well known, the doubly differential cross section in the two hemisphere
masses factorizes in the limit that both masses are small~\cite{hep-ph/0703207,0709.2709,0803.0342,1005.1644,1105.3676,1105.4628},
\begin{equation}
  \frac{d^2 \sigma}{d m_L d m_R} = H \left( Q, \mu \right) \int d k_L d k_R J
  ( m_L^2 - k_L Q, \mu ) J ( m_R^2 - k_R Q, \mu ) S
  ( k_L, k_R, \mu ) \label{factform}
\end{equation}
The $\mu$ dependence of all these functions is known to 3 loops at fixed order
and has been resummed to the next-to-next-to-next-to-leading logarithmic level
(N$^3$LL). This resummation only accounts for the global logarithms.
For some observables, such as thrust $T$ all the logs are global. 
For thrust, when  $\tau = 1 - T
\approx 0$ then $\tau \approx \frac{1}{Q^2} ( m_L^2 + m_R^2
)$ and Eq.~\eqref{factform} reduces to~\cite{0803.0342}
\begin{equation}
  \frac{d \sigma}{d \tau} = H ( Q, \mu ) \int d k  J ( \tau
  - \frac{k}{Q}, \mu ) J ( \tau - \frac{k}{Q}, \mu ) S_T
  ( k, \mu ) \label{thrustfact}
\end{equation}
with $S_T ( k, \mu ) = \int d k_L d k_R S ( k_L, k_R, \mu )\delta(k-k_L-k_R)$. In this case, each
function has only one scale and all the logs can be resummed. In contrast,
 if there are multiple scales, like $m_R$, $m_L$ and $Q$, one cannot
resum all the large logarithms so simply. To see the difficulty more clearly,
we can write the soft function as
\begin{equation}
  S ( k_L, k_R, \mu ) = S_{\mu} ( \ln \frac{k_L}{\mu} )
  S_{\mu} ( \ln \frac{k_R}{\mu} ) S_f ( \ln \frac{k_L}{k_R})
\end{equation}
Because $S_{\mu} ( L )$ depends on $\mu$, its large logarithms can
be resummed using the renormalization group. $S_f ( L )$ on the
other hand is some finite function whose resummation is more subtle. The
non-global logarithms are those contained in $S_f ( L )$. Note that
thrust is only sensitive to $S_f ( 0 )$, so these non-global
logarithms do not inhibit resummation of logs of thrust.

There are no double logarithms in $S_f ( L )$. Instead $S_f$ has
single logarithms, of the form $( \alpha_s L )^n$ and subleading
logarithms, of the form $ \alpha_s^m L^n$ with $m > n$, both starting from $\alpha_s^2$. The
coefficient of the 2-loop leading non-global logarithm was computed in Ref.~\cite{Dasgupta:2001sh}, where it
was found $\left. S_f ( L )\right|_{\mathrm{leading}} = -\left( \frac{\alpha_s}{2 \pi} \right)^2 L^2
\left( C_F C_A \frac{\pi^2}{3} \right)$. The complete form of $S_f ( L)$ at 2 loops was computed in Refs.~\cite{1105.3676,1105.4628}, revealing subleading logarithms in
both the $C_F C_A$ and $n_f T_F C_F$ color structures, as well non-singular
pieces. When $L$ is large but $\alpha_s^2 L$ is small, the leading non-global
logarithms dominate.  Unfortunately, resumming the leading logarithms is not
as simple as writing $S_f ( L ) = \exp \left[ -\left(
\frac{\alpha_s}{2 \pi} \right)^2 L^2 \left( C_F C_A \frac{\pi^2}{3} \right)
\right]$. Although the non-global logs do exponentiate in this way, due to
non-Abelian exponentiation, at each order in perturbation theory, new
maximally non-Abelian color structures appear which also scale like 
$(\alpha_s L )^n$. For example, at 3 loops, as we will see, there is a
term $\sim ( \alpha_s L )^3 C_F C_A^2$ which is not contained in the
exponentiated 2-loop result.

A number of simplifications facilitate the extraction of the leading
non-global logarithm to high orders. First, one can consider a simpler
observable, the right-hemisphere mass. By integrating inclusively over the
left hemisphere, logs of $\frac{m_R}{m_L}$ are replaced by logs of
$\frac{m_R}{Q}$. In particular, the coefficients of these logs are
exactly the same in the left-right hemisphere case and the right
hemisphere case. In addition, removing the restriction $m_L \ll Q$ probes the
hard multijet region, which is outside of the validity of the factorization
formula in Eq. (\ref{factform}). This is ordinarily dangerous: the
left-hemisphere integral contributes something proportional to $\alpha_s$ with
no logarithm, which may multiplying $( \alpha_s \ln^2 m_R )^n$ terms
from the right-hemisphere integral. However, these logarithms are
global in nature and can be  resummed~\cite{1105.4628}.

We can go further than taking $m_L = Q$, we can take $Q \rightarrow \infty$.
This introduces UV divergences. Regulating them in dimensional regularization
and $\overline{\text{MS}}$, the scale $m_R$ in the original logarithm gets
replaced by $\mu$. To be more concrete, we can define the right-hemisphere
soft function as
\begin{equation}
  S_R ( k, \mu ) = \sum_{m = 0}^{\infty} \int d \Pi_m |
  \langle k_1 \cdots k_m | Y_n^{\dag} Y_{\nbar} | 0 \rangle
  |^2 \delta \left( k - \sum_i n \cdot k_i \theta_R ( k_i ) \right) \label{RHsoft},
\end{equation}
where $\Pi_m$ denotes phase space of $m$ soft parton emission, and
$Y_n$ and $Y_{\nbar}$ are fundamental Wilson lines stretching from collision point, at the origin, to infinity.
This function is infrared finite, but UV divergent. The UV divergences are
removed in $\overline{\text{MS}}$, generating the $\mu$-dependence. We can
write it as
\begin{equation}
  S_R ( k, \mu ) = S_{\mu} ( \ln \frac{k}{\mu} ) S_{R f}( \ln \frac{k}{\mu} )
  \label{Smu}
\end{equation}
with $S_{\mu}(L)$ the same function, containing the global
logarithms, as in the double-hemisphere soft function. $S_{\mu}(L)$ is constrained by renormalization-group invariance of the factorization formula in Eq.
(\ref{thrustfact}) to be related to the hard and jet functions. We do not have
a factorization formula containing $S_R \left( k, \mu \right)$. We can
nevertheless differentiate with respect to $\mu$ to find an RGE for $S_R \left(
k, \mu \right)$. This RGE will be local in Laplace space. Defining
\begin{equation}
  \tilde{s}_R \left( \nu, \mu \right) = \int d k e^{- \nu k} S_R \left( k, \mu
  \right)
\end{equation}
we have
\begin{equation}
  \frac{\partial}{\partial \ln \mu} \tilde{S}_R \left( \nu, \mu \right) =
  \Gamma_R ( \ln \mu \nu ) \tilde{S}_R \left( \nu, \mu \right)
\end{equation}
A similar equation holds for the thrust soft function. In that case, the
anomalous dimension $\Gamma ( L )$ is linear in $L$ to all orders
in $\alpha_s$: $\Gamma_{\text{hemi}} ( L ) \sim
\Gamma_{\text{cusp}} ( \alpha_s ) L + \gamma_{\text{reg}} (
\alpha_s)$. For the single-hemisphere soft function $\Gamma_R ( L)$
 has global terms linear in $L$ which are proportional to the cusp
anomalous dimension, from $S_{\mu} \left( L \right)$. As we will see, it also
has nonlinear terms corresponding to the non-global logarithms. Although the
right-hemisphere soft function has no all-orders relation to the hemisphere
soft function, their leading non-global logarithms will agree.

The next important observation is that the leading logarithms are entirely
generated by regions of real and virtual phase space which are strongly
ordered in energy~\cite{Dasgupta:2001sh}. A region where two particles' momenta are comparable will
contribute a finite amount, but not a large logarithm. In addition, when two
gluons are present in the right hemisphere and $E_1 \gg E_2$, then only $E_1$
contributes to the hemisphere mass. Thus, $E_2$ contributes only a finite
correction and this configuration cannot give a leading logarithm. Therefore,
at order $\alpha_s^n$, the right-hemisphere mass distribution can be
calculated from the cross section for producing $n$ strongly ordered gluons,
with exactly one going into the right hemisphere. There are also virtual
contributions, and real-virtual contributions. But in each case, only one real
emission can go into the right hemisphere, as will be clear below.

\section{Strong energy ordering}
\label{sec:seo}
In this section, we review the structure of the real, virtual and real-virtual
integrands relevant for the leading non-global logarithm at large
$N_c$ limit~\cite{Bassetto:1984ik}. 
While simplifications arising from the strong-energy-ordering (SEO) limit have been
known for decades, we try to provide more explicit details than we have found in the literature.
Hopefully, our exposition will clarify the set of approximations going into the NGL calculation.
A reader already familiar with SEO can skip this section.

\subsection{Real emission}
 To begin, consider the cross section for emission of $m$ gluons
off classical quark sources in the $a^{\mu}$ and $b^{\mu}$ directions. The
differential cross section for real-emission is then
\begin{equation}
  \frac{1}{\sigma_0} d \sigma_m = \frac{1}{m!}d \Phi_m \left| \mathcal{M}_{a b}^{1 \cdots
  m} \right|^2
\end{equation}
where $\sigma_0$ is the tree-level cross section and the phase space is
\begin{equation}
  d \Phi_m = \prod_{i = 1}^m \frac{d^3 p_i}{\left( 2 \pi \right)^3 2 \omega_i}
  = \prod_{i = 1}^m \frac{\omega_i d \omega_i}{4 \pi^2} \frac{d \Omega_i}{4
  \pi}
\end{equation}

In the limit that the energy of the gluons is strongly ordered, at large $N_c$
the matrix-element squared can be written as~\cite{Bassetto:1984ik}
\begin{equation}
  \left| \mathcal{M}_{a b}^{1 \cdots m} \right|^2 = \left| \langle p_1 \cdots
  p_m \left| Y_a^{\dag} Y_b \right| 0 \rangle \right|^2 =  N_c^m
  g^{2 m} \sum_{\text{perms of } 1 \cdots m} \frac{\left( p_a \cdot p_b
  \right)}{\left( p_a \cdot p_1 \right) \left( p_1 \cdot p_2 \right) \cdots
  \left( p_m \cdot p_b \right)} \label{generalM}
\end{equation}
It does not matter if $E_1 \gg E_2 \gg \cdots \gg E_m$ or if the gluons are ordered
in some other permutation; because they are identical particles, the matrix
element is independent of the gluon labels. 

To simplify cross section formula, it is helpful to pull out the energies from
the dot-products, by  writing
\begin{equation}
  ( i j ) \equiv \frac{p_i \cdot p_j}{\omega_i \omega_j} 
  = 1 - \cos
  \theta_{i j} 
\label{roundbracket}
\end{equation}
where $\theta_{i j}$ is the angle between the directions $\vec{p}_i$ and
$\vec{p}_j$. Then we define the radiator function as
\begin{equation}
  {\cW}_{a b}^{1 \cdots m} = \frac{( a b )}{( a 1) ( 12 ) \cdots ( m b )} \label{Wdef}
\end{equation}
and
\begin{equation}
  \mathcal{P}_{a b}^{1 \cdots m} =  \sum_{\text{perms of}~1
  \cdots m} \cW_{a b}^{1 \cdots m} \label{Pdef}
\end{equation}
so that
\begin{equation}
  \left| \mathcal{M}_{a b}^{1 \cdots m} \right|^2 =  N_c^m g^{2 m}
  \frac{1}{\omega_1^2 \cdots \omega_m^2} \mathcal{P}_{a b}^{1 \cdots m}
  \label{M2form}
\end{equation}
We thus write
\begin{equation}
  \frac{1}{\sigma_0} d \sigma_m = \sum_m \frac{1}{m!}\bar{\alpha}^m \prod_{i = 1}^m
  \frac{d \omega_i}{\omega_i} \frac{d \Omega_i}{4 \pi} \mathcal{P}_{a b}^{1
  \cdots m} \label{sigom}
\end{equation}
with
\begin{equation}
  \bar{\alpha} \equiv N_c \frac{\alpha_s}{\pi}
\end{equation}

It is easy to understand the form of Eq. (\ref{generalM}) or Eq.
(\ref{M2form}). For one-emission $\left| \mathcal{M}_{a b}^1 \right|^2$ is
just the eikonal vertex summed over polarizations~\cite{Dokshitzer:1991wu}
\begin{align}
  \left| \mathcal{M}_{a b}^1 \right|^2 &
 = g^2 \left| \text{tr}
  \left( T^a \frac{p_a^{\mu}}{p_a \cdot p_1} + T^a_{} \frac{p_b^{\mu}}{p_b
  \cdot p_1} \right) \left( T^{b\dag} \frac{p_b^{\nu}}{p_b \cdot p_1} +
  T^{b\dag} \frac{p_b^{\nu}}{p_b \cdot p_1} \right) \left( - g^{\mu \nu}
  \delta^{a b} \right) \right|
\\
  &=2 g^2 C_F \frac{\left( p_a \cdot p_b \right)}{\left( p_a \cdot p_1 \right)
  \left( p_1 \cdot p_b \right)}
\\
  & =  N_c g^2 \frac{1}{\omega_1^2} \mathcal{P}_{a b}^1 +
  \mathcal{O}(N_c^0 ).
\end{align}
Here strong-ordering only goes into the use of the eikonal approximation.

For two gluons, suppose first that $\omega_1 \gg \omega_2$. Then we can think
of the quarks (Wilson lines) as emitting gluon 1 first, with a rate
proportional to $\mathcal{P}_{a b}^1$. Since gluon 2 is much softer, it views
gluon 1 as a source for radiation; thus we have a new adjoint Wilson line the
$1$ direction. At large $N_c$ this Wilson line is equivalent to a fundamental
Wilson line which forms a dipole with the $a$ antiquark and an
antifundamental Wilson line which forms a dipole with the $b$ quark. These two
dipoles then radiate proportional to $\mathcal{P}_{a 1}^2$ and $\mathcal{P}_{1
b}^2$ respectively. So we have
\begin{equation}
  \mathcal{P}_{a b}^{12} = \mathcal{P}_{a b}^1 \left[ \mathcal{P}_{a 1}^2 +
  \mathcal{P}_{1 b}^2 \right] \label{Pab1}
\end{equation}
which is easy to check using Eqs. (\ref{Wdef}) and (\ref{Pdef}). It is also
true that
\begin{equation}
  \mathcal{P}_{a b}^{12} = \mathcal{P}_{a b}^2 \left[ \mathcal{P}_{a 2}^1 +
  \mathcal{P}_{2 b}^1 \right] \label{Pab2}
\end{equation}
which can be understood by repeating the above argument when $\omega_2 \gg
\omega_1$.

For $m$ emissions, one continues this recursive picture of Wilson lines
begetting new Wilson lines. Some combinatorics then establishes the general
result in Eq. (\ref{generalM}). This SEO dipole picture is of course
well-known and critical to the success of Monte Carlo event generators and
many QCD calculations.

The equivalence of Eqs. (\ref{Pab1}) and (\ref{Pab2}) is guaranteed since the
gluons are identical. A useful set of related identities for the radiator
function are
\begin{equation}
  \cW_{a b}^{12 \cdots m} =   \cW_{a r}^{1 \cdots
  \left( r - 1 \right)} \cW_{a b}^r \cW_{r b}^{\left( r + 1
  \right) \cdots m}, \hspace{1em} 1 \leqslant r \leqslant m \label{Widen}
\end{equation}
For example, $\cW_{a b}^{12} =  \cW_{a b}^1
\cW_{1 b}^2 = \cW_{a 2}^1 \cW_{a b}^2$ and
$\cW_{a b}^{123} =  \cW_{a b}^1 \cW_{a
b}^{23} = \cW_{a 2}^1 \cW_{a b}^2 \cW_{2 b}^3 =
\cW_{a 3}^{12} \cW_{\text{ab}}^3$.

\subsection{Virtual and real-virtual corrections}
Virtual contributions to cross sections can also be understood in the
SEO limit~\cite{Bassetto:1984ik,Bassetto:1982ma}.
For virtual momenta, we should think of the SEO approximation as not just
energy ordering but ordering of all the components of the momenta. For
on-shell momenta, having small energy implies all the components are small.
For off-shell momenta, this is not true. However in the region of virtual
phase space where the energy is small but the momentum is large the virtual
momentum is highly off-shell. This off-shell region can contribute finite
parts to a cross section, but cannot contribute to the leading logarithms.
Thus, in the relevant region of phase space, the virtual momenta is nearly on
shell and can be treated like a real emission.

\begin{figure}[t]
\begin{center}
\includegraphics[width=0.32\textwidth]{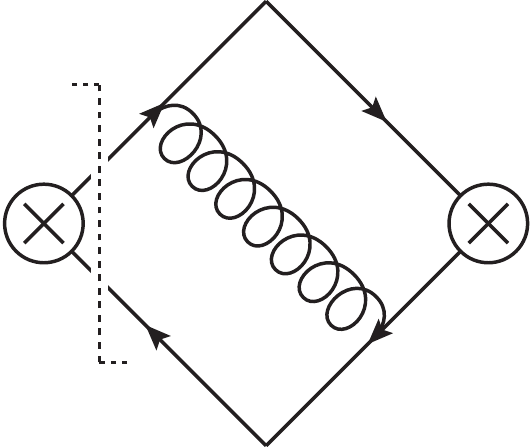}
\hspace{1.5cm}
\includegraphics[width=0.3\textwidth]{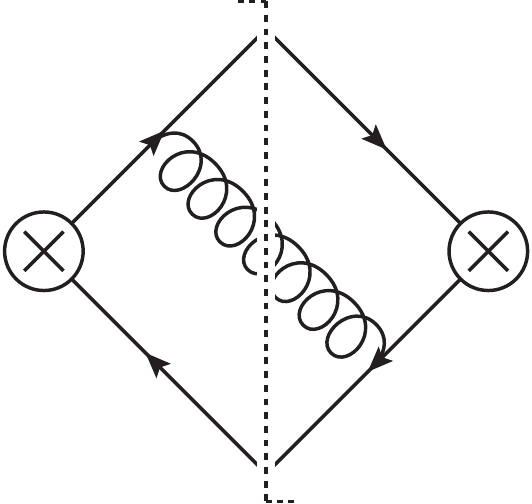}
\end{center}
\caption{Both the virtual contribution (left) and real contribution (right) to the $\als$ cross section can be drawn as cut diagrams with the same topology.}
  \label{fig:cuts}
\end{figure}

At order $\alpha_s$, the virtual contribution to a cross section contributes by interfering with the tree-level graph.
This interference can be drawn
as a cut diagram (Fig.~\ref{fig:cuts}, left), which is nearly
identical to the real emission graph (Fig.~\ref{fig:cuts}, right). By moving the cut, the propagator in the virtual graph is replaced by an 
on-shell condition. That is,
\begin{equation}
d \sigma_V \propto - \mathcal{P}_{a b}^1 \frac{d^4 p_1}{(2 \pi)^4} \frac{i}{p^2 + i \varepsilon}
,\quad\quad
  d \sigma_R \propto \mathcal{P}_{a b}^1 \frac{d^4 p_1}{( 2 \pi)^3} \theta ( \omega_1 ) \delta ( p_1^2 ),  \hspace{1em} 
\end{equation}
with the minus sign in the virtual contribution arising since the cut is on
one side of the emissions only. To relate the two contributions, we can cut the propagator ourselves through the identity
\begin{equation}
  \frac{i}{p^2 + i \varepsilon} = \text{PV} \left\{ \dfrac{i}{p^2} \right\} +
  \pi \delta ( p^2 ) \theta ( p_0 ) + \pi \delta (  p^2 ) \theta ( - p_0 )
\label{propdec}
\end{equation}
Since all the components of $p_1^{\mu}$ are very small, by the SEO assumption, the
virtual gluon is nearly on shell. In this limit, the
principal value contribution vanishes. More generally, since the principle value is not IR-sensitive, it can contribute only a finite part to the cross
section, not a large logarithm. The $\delta ( p^2)$ terms in Eq.~\eqref{propdec} have
support for either $p_0 > 0$ or $p_0 < 0$, each of which give the same contribution.
Thus we can replace $\frac{i}{p^2 + i \varepsilon} \rightarrow 2 \pi \theta( p_0 ) \delta ( p^2 )$ in the virtual contribution
showing it to have the same form as the real emission, up to a sign.

Another way to understand the connection between real and virtual is to use
that in a sufficiently inclusive cross section, the large logarithm from real emission
must be exactly canceled by virtual corrections. Thus we should be able to
represent the virtual contributions as integrals over momenta of exactly the
same form as the real emissions. For example, at 1 loop, we would have
\begin{equation}
  d \sigma_V = - d \sigma_R \propto - N g^2 \frac{1}{\omega_1^2}
  \mathcal{P}_{a b}^1
\end{equation}
Let us abbreviate this with
\begin{equation}
  \cW_R = \mathcal{P}_{a b}^1, \hspace{1em} \cW_V = -
  \cW_R \label{WRandV}
\end{equation}
For an observable which is not totally inclusive, like the hemisphere mass,
there will be an incomplete cancellation between the real and virtual
corrections, leaving a large logarithm.

For more general notation, let us write $\cW_{R V \cdots R}$ to
indicate that the hardest gluon, $1$, is real, the second hardest, 2, is
virtual and so on down to the softest, $m$, which in this case is real.
Thus, for example, at $\alpha_s^2$, the differential cross section can be
written as, using Eq. (\ref{sigom})
\begin{multline}
  \frac{1}{\sigma_0} d \sigma_m = 
\hspace{5mm} \bar{\alpha} \frac{d \omega_1}{\omega_1} \frac{d
  \Omega_1}{4 \pi} \left( \cW_R + \cW_V \right)
 \\
  + \frac{\bar{\alpha}^2}{2!} \frac{d \omega_1}{\omega_1} \frac{d \Omega_1}{4 \pi}
  \frac{d \omega_2}{\omega_2} \frac{d \Omega_2}{4 \pi} \left( \cW_{R
  R} + \cW_{R V} + \cW_{V R} + \cW_{V V} \right)
\\
  + \frac{\bar{\alpha}^3}{3!} \frac{d \omega_1}{\omega_1} \frac{d \Omega_1}{4 \pi}
  \frac{d \omega_2}{\omega_2} \frac{d \Omega_2}{4 \pi} \frac{d
  \omega_3}{\omega_3} \frac{d \Omega_3}{4 \pi} \left( \cW_{R R R} +
  \cW_{R R V} + \cdots \right)
\end{multline}

For two emissions, the gluons can be either real or virtual. If both are real,
we get the expression in Eq. (\ref{Pab1}):
\begin{equation}
  \cW_{R R} = \mathcal{P}_{a b}^{12} = \mathcal{P}_{a b}^1 \left[
  \mathcal{P}_{a 1}^2 + \mathcal{P}_{1 b}^2 \right]
\end{equation}
This holds for either $\omega_1 \gg \omega_2$ or $\omega_2 \gg \omega_1$. 
If the harder gluon
is real and the softer gluon is virtual, the real emission establishes the
$( a 1 )$ and $( 1 b )$ dipoles, which then each
contribute a virtual contribution. So we have
\begin{equation}
  \cW_{R V} = - \mathcal{P}_{a b}^1 \left[ \mathcal{P}_{a 1}^2 +
  \mathcal{P}_{1 b}^2 \right] 
\end{equation}
On the other hand, if the harder gluon ($1$) is virtual, then the virtual
graph does not produce any new dipoles. So we get $-
\mathcal{P}_{a b}^1$ for the first emission, but have only the original $a b$
dipole to produce subsequent emissions. This dipole then produces the real
emission and we have
\begin{equation}
  \cW_{V R} = - \mathcal{P}_{a b}^1 \mathcal{P}_{a b}^2
\end{equation}
If both the harder and softer gluon are virtual, then the $a b$ dipole
produces both, each get a minus sign, and we find
\begin{equation}
  \cW_{V V} = \mathcal{P}_{a b}^1 \mathcal{P}_{a b}^2
\end{equation}
Thus there are 2 independent integrands
\begin{align}
  B_1 &= \mathcal{P}_{a b}^{12} = \cW_{R R} = - \cW_{R V}\\
  B_2 &= \mathcal{P}_{a b}^1 \mathcal{P}_{a b}^2 = \cW_{V V} = -
  \cW_{V R}
\end{align}

At order $\alpha_s^3$, we construct the real and virtual integrands in the
same iterative way. For example, the contribution with all gluons virtual
comes from 3 uncorrelated emissions from the $(a b)$ dipole, each with a minus
sign:
\begin{equation}
  \cW_{V V V} = - \cW_{a b}^1 \cW_{a b}^2
  \cW_{a b}^3
\end{equation}
When the hardest gluon is real and the second is virtual, we get $\mathcal{-
P}_{a b}^{12}$ as above. Since the second gluon is virtual, it does not
produce a new dipole, so the 3rd emission comes from the $\left( a 1 \right)$
and $\left( 2 b \right)$ dipoles only. Thus we find
\begin{equation}
  \cW_{R V R} = - \cW_{R V V} = - \mathcal{P}_{a b}^{12}
  \left( \mathcal{P}_{a 1}^3 + \mathcal{P}_{b 1}^3 \right)
\end{equation}
In total at order $\alpha_s^3$, there are 4 independent integrands:
\begin{align}
  C_1 &= \mathcal{P}_{a b}^{123} = \cW_{R R R} = - \cW_{R R V}
  \label{Cs}\\
  C_2 &= \mathcal{P}_{a b}^{12} \left( \mathcal{P}_{a 1}^3 + \mathcal{P}_{b
  1}^3 \right) = \cW_{R V V} = - \cW_{R V R}
\\
  C_3 &= \mathcal{P}_{a b}^1 \mathcal{P}_{a b}^{23} = - \cW_{V R R} =
  \cW_{V R V}
\\
  C_4 &=\mathcal{P}_{a b}^1 \mathcal{P}_{a b}^2 \mathcal{P}_{a b}^3 =
  \cW_{V V R} = - \cW_{V V V}
\end{align}
Summing all eight contributions gives zero, as expected since there can be
no large logarithms in an inclusive cross section. The procedure for
constructing the real and virtual contributions to $\left| \mathcal{M}
\right|^2$ to arbitrary order should now be clear by generalizing these
examples.

\section{The non-global hemisphere mass integral}
\label{sec:nghmi}

The procedure defined in the previous sections provide the real and
virtual contributions to $| \mathcal{M}|^2$ in the SEO approximation. To construct an
observable, we have to integrate these matrix elements against a measurement
function. Since virtual gluons are never measured, this function is only
sensitive to the gluons which are real. In this section, we work out the integrand at up to 3 loops and outline
the procedure for higher loops. Above 3 loops, we find it simpler to extract the  NGL integrand 
using the BMS equation, as explained in the next section.

To avoid dealing with distributions, we work with the cumulant
right-hemisphere mass defined as $\rho = \frac{M_R}{Q}$. We then have
\begin{equation}
  S ( \rho ) = \frac{1}{\sigma_0} \int_0^Q d m_L  \int_0^{\rho Q}
  d m_R \frac{d^2 \sigma}{d m_L d m_R} 
\end{equation}
We can therefore write the cross section as the integral of the matrix-element
squared times a measurement function
\begin{equation}
  S ( \rho ) = \int d \Phi_m \left| \mathcal{M}_{a b}^{1 \cdots m}
  \right|^2 u ( \{ p_i \} )
\end{equation}
where the measurement function for the hemisphere mass cumulant at leading power is
\begin{equation}
  u ( \{p_i \} ) = \Theta \Big( \rho Q - \sum_i  2 ( p_i \cdot n) \theta_R (p_i ) \Big)
\end{equation}
Working in a frame where the jet are back-to-back in the $n^{\mu} = ( 1,
\vec{n} )$ and $\nbar^{\mu} = ( 1, - \vec{n} )$ directions,
the right-hemisphere projector is $\theta_R ( p ) \equiv \theta
( \vec{p} \cdot \vec{n} )$. Similarly, the left-hemisphere
projector is $\theta_L ( p ) \equiv \theta ( - \vec{p} \cdot
\vec{n} ) = 1 - \theta_R ( p )$. Since only the hardest gluon
in the hemisphere will contribute, we can equally well use
\begin{equation}
  u ( \{p_i \} ) = \prod_{p_i} u ( p_i )
\label{uprod}
\end{equation}
where
\begin{equation}
  u \left( p \right) = \Theta ( \rho Q - 2 p \cdot n ) \theta_R
  ( p ) + \theta_L ( p )
\end{equation}
That we can treat the emissions independently greatly simplifies the
calculation.\footnote{The measurement function factorizes into a product of terms exactly when transformed into Laplace space. For the
leading NGLs, which we consider here, Eq.~\eqref{uprod} is enough.}

For one emission, we can write the cumulant as
\begin{equation}
  S^{( 1 )} ( \rho ) = \ba\int \cW_R u ( p_1) +\ba \int \cW_V,
\end{equation}
where no phase space constraint is imposed on the virtual gluon, as
the measurement operator does not act on the virtual gluons. Let us
write more suggestively,
\begin{equation}
  u ( p_1 ) = \Theta ( \rho Q - 2 p_1 \cdot n ) \theta_R( p_1 ) = 1_R \theta_{1 < \rho} + 1_L
\end{equation}
$1_R$ means that gluon 1 goes to the right and $\theta_{1 < \rho}$ means that
gluon 1's contribution to the hemisphere mass is not larger than $\rho$. Using
Eq. (\ref{WRandV}) the $\mathcal{O} ( \alpha_s )$ result is then
\begin{equation}
  S^{( 1 )} ( \rho ) =\ba \int \mathcal{P}_{a b}^1 (
  1_R \theta_{1 < \rho} + 1_L - 1 ) = - \ba\int \mathcal{P}_{a b}^1 1_R
  \theta_{\rho < 1}
\end{equation}
This is the global logarithm. To all orders, the global logarithm is given by
the exponentiation of this term.

For two emissions, we have
\begin{multline}
  S^{\left( 2 \right)} \left( \rho \right) = \ba^2\int_{E_1>E_2} \left( 1_R \theta_{1 < \rho}
  + 1_L \right) \left( 2_R \theta_{2 < \rho} + 2_L \right) \cW_{R R} +
 \ba^2 \int_{E_1>E_2} \left( 1_R \theta_{1 < \rho} + 1_L \right) \cW_{R V}
\\
  + \ba^2\int_{E_1>E_2} \left( 2_R \theta_{2 < \rho} + 2_L \right) \cW_{V R} + \ba^2\int_{E_1>E_2}
  \cW_{V V}
\end{multline}
Our notation is defined so that gluon 1 is much harder than gluon 2. Therefore
$\theta_{\rho < 1} \theta_{\rho < 2} = \theta_{\rho < 2}$. For the
same reason, we can drop the symmetry factor $1/2!$, since
only one energy ordering is picked out for the phase space integral. We can then write
$S^{\left( 2 \right)} ( \rho )$ suggestively as
\begin{equation}
  S^{( 2 )} ( \rho ) =\ba^2 \int_{E_1>E_2} 1_R 2_R \theta_{\rho < 1}
  \theta_{\rho < 2} \left( \mathcal{P}_{a b}^1 \mathcal{P}_{a b}^2 \right) -
\ba^2  \int_{E_1>E_2} 1_L 2_R \theta_{\rho < 2} \left( \mathcal{P}_{a b}^{12} -
  \mathcal{P}_{a b}^1 \mathcal{P}_{a b}^2 \right) \label{2simpl}
\end{equation}
The first term here is the global contribution with both gluons going right
but uncorrelated. \ If we average the first integral over the same thing with
$E_2 > E_1$, we can drop the energy ordering and have simply
\begin{equation}
\ba^2  \int_{E_1 > E_2} 1_R 2_R \theta_{\rho < 2} \left( \mathcal{P}_{a b}^1
  \mathcal{P}_{a b}^2 \right) =  \frac{\ba^2}{2} \int 1_R 2_R \theta_{\rho < 1}
  \theta_{\rho < 2} \mathcal{P}_{a b}^1 \mathcal{P}_{a b}^2 = \frac{1}{2}
  \left[ S^{\left( 1 \right)} \left( \rho \right) \right]^2,
\end{equation}
which agrees with the second-order expansion of $\exp \left( S^{\left( 1
\right)} \left( \rho \right) \right)$. The second term in Eq. (\ref{2simpl})
when integrated gives the leading non-global logarithm. Explicitly,
\begin{align}
  S^{( 2 )}_{\text{NG}} = & - \bar{\alpha}^2 \int_{E_1>E_2} 1_L 2_R \theta_{\rho
  < 2} \left( \mathcal{P}_{a b}^{12} - \cW_{a b}^1 \cW_{a b}^2
  \right) 
  \nn
  \\
  =&- \bar{\alpha}^2 \int_0^Q \frac{d \omega_2}{\omega_2} \int_{\text{right}}\frac{d \Omega_2}{4
  \pi} \int_{\omega_2}^{Q} \frac{d \omega_1}{\omega_1} \int_{\text{left}}\frac{d \Omega_1}{4 \pi} 
\theta \left( \omega_2 -
  \frac{\rho Q}{2 ( n 2 )} \right)
\nn
\\
&
\hspace{30mm}
  \times \left[ \frac{( n \nbar )}{( n 1 ) ( 12 )
  ( 2 \nbar )} +  \frac{( n \nbar )}{( n 2 ) ( 21 )
  ( 1 \nbar )} - \frac{( n \nbar )}{( n 1
  ) ( 1 \nbar )} \frac{( n \nbar )}{( n 2
  ) ( 2 \nbar )} \right]   \nn \\
  = &- \bar{\alpha}^2 \frac{\pi^2}{24} \ln^2 \rho + \text{less singular terms}
  \label{2loopform}
\end{align}
This integrand is exactly that given by Eq. (8) of Ref.~\cite{Dasgupta:2001sh}.

A simplifying observation is that because the non-global integral has no
collinear singularities, we can replace the $\theta$ function on
hemisphere mass with a
simpler one on energy: $\theta \left( \omega_2 - \frac{\rho Q}{2( n 2)} \right) \rightarrow \theta ( \omega_2 - \rho )$. The
difference produces only subleading terms. This is not allowed in the global
logarithmic terms because unregulated collinear divergences would arise, but is allowed for non-global ones.

The first new result here is the 3-loop integrand. Following the same
procedure outlined above, we find
\begin{multline}
  S^{\left( 3 \right)} \left( \rho \right) = \ba^3 \int_{E_1>E_2>E_3} 1_R 2_R 3_R \theta_{\rho <
  3} \left( - C_4 \right) + \ba^3\int_{E_1>E_2>E_3} 1_R 2_L 3_R \theta_{\rho < 3} \left( C_3 -
  C_4 \right)
\\
  +\ba^3 \int_{E_1>E_2>E_3} 1_L 2_R 3_R \theta_{\rho < 3} \left( C_2 - C_4 \right) +\ba^3 \int_{E_1>E_2>E_3} 1_L 2_L
  3_R \theta_{\rho < 3} \left( - C_1 + C_2 + C_3 - C_4 \right)
\end{multline}
with $C_1$, $C_2$, $C_3$ and $C_4$ given in Eqs. (\ref{Cs}). To find the
3-loop NGLs, we have to remove the global logarithms. To find the purely 3-loop contribution,
we should also remove the exponentiation of all the previous
terms. Writing the cumulant in an exponentiated form,
\begin{equation}
  S = \exp \left( S^{\left( 1 \right)} + S_{\text{NG}}^{( 2 )} +
  S_{\text{NG}}^{\left( 3 \right)} + \cdots \right),
\end{equation}
we find
\begin{equation}
  S_{\text{NG}}^{\left( 3 \right)} = \ba^3 \int_{E_1>E_2>E_3} 1_L 2_R 3_R \theta_{\rho < 3} \left(
  \overline{C_2 - C_4} \right) + \ba^3 \int_{E_1>E_2>E_3} 1_L 2_L 3_R \theta_{\rho < 3} \left( -
  C_1 + C_2 + C_3 - C_4 \right) \label{SNG3}
\end{equation}
where
\begin{equation}
  \overline{C_2 - C_4} = \mathcal{P}_{a b}^{12} \left( \mathcal{P}_{a 1}^3 +
  \mathcal{P}_{1 b}^3 \right) - \mathcal{P}_{a b}^1 \mathcal{P}_{a b}^2
  \mathcal{P}_{a b}^3 - \mathcal{P}_{a b}^2 \left[ \mathcal{P}_{a b}^{13} -
  \mathcal{P}_{a b}^1 \mathcal{P}_{a b}^3 \right] - \mathcal{P}_{a b}^3 \left[
  \mathcal{P}_{a b}^{12} - \mathcal{P}_{a b}^1 \mathcal{P}_{a b}^2 \right]
\end{equation}
Both terms in Eq. (\ref{SNG3}) are free of collinear singularities.

Note that at 3 loops, it is not true that only the softest gluon goes to the
right. The $2_R$ in Eq.~\eqref{SNG3} indicates that a contribution also comes from the middle gluon going to the right. 
However, this is somewhat of an illusion. Since all the integrands which give $\pm C_2$ or $\pm C_4$ have
the  second gluon virtual, all the $2_R$ terms in the first term in Eq.
(\ref{SNG3}) correspond to virtual emissions. Thus, while there is a $2_R$ contribution, a real gluon which is
not the softest never actually goes into the right hemisphere.

While we could proceed to evaluate $S_{\text{NG}}^{\left( 3 \right)}$ at this
point, it is somewhat simpler first to reproduce $S_{\text{NG}}^{\left( 3
\right)}$ from a recursive formula (the BMS equation). This will simplify the
calculation at 3 loops and beyond, as we will now see.

\section{BMS equation}\label{sec:BMS}

The BMS equation~\cite{Banfi:2002hw} is an integro-differential equation whose solution gives the
leading NGLs (those of the form $(\bar{\alpha} \ln \rho)^n$, for $n>1$) at large $N_c$. 
The derivation of the BMS equation for the hemisphere mass proceeds
identically to the derivation for the out-of-jet energy given in Ref.~\cite{Banfi:2002hw}.
We have nothing profound to add to the derivation, so we simply present the result. For the
hemisphere case, the BMS equation becomes
\begin{equation}
  \partial_\tL G_{a b} ( L ) = \int  \frac{d \Omega_j}{4 \pi} 
  \cW_{a b}^j \left[ \theta_L ( j ) G_{a j} ( L) G_{j b} ( L ) - G_{a b} ( L ) \right],
  \label{BMSbig}
\end{equation}
Here, $\cW_{ab}^j$ is the dipole radiator, from Eq.~\eqref{Wdef}:
\begin{equation}
\cW_{ab}^j = \frac{(ab)}{(aj)(jb)}
\end{equation}
with $(ab) = \frac{a \cdot b}{\omega_a \omega_b}$ 
and $\theta_L ( j )$ restricts the angular integral to being over the left-hemisphere. Recall that our convention is such that $n^\mu$ points to the right hemisphere and $\nbar^\mu$ to the left hemisphere
 and that
\begin{equation}
  \cos\theta_n = -1, \qquad \cos\theta_{\nbar} = 1
\end{equation}
The solution to the BMS equation are a set of functions  $G_{a b} ( L )$ indexed by lightlike directions
$a^\mu$ and $b^\mu$ (equivalently angles $\Omega_a$ and $\Omega_b$ on the 2-sphere).   These functions, when evaluated at
\begin{equation}
  L= \hL \equiv  N_c \frac{\als}{\pi} \ln \frac{1}{\rho},
\end{equation}
give all the single (global and non-global)
logarithms of the hemisphere mass from a color dipole in $a^\mu$ and $b^\mu$ directions. In
particular, the hemisphere mass NGLs are in $G_{n\nbar}(\hL)$.
There are additional single logarithms coming from the 1-loop running of $\als$. These can be easily included~\cite{Dasgupta:2001sh}, so we simply ignore them for simplicity.

To extract just the NGLs, following BMS we write
\begin{equation}
  G_{a b} ( L ) = g_{a b} ( L ) \exp \left( - \tL \int_{\text{right}} 
  \frac{d \Omega_j}{4 \pi} \cW_{a b}^j \right),
\end{equation}
which leads to
\begin{equation}
  \partial_\tL g_{a b} ( L ) = \int_{\text{left}}  \frac{d
  \Omega_j}{4 \pi}  \cW_{a b}^j \left[ U_{a b j} ( L ) g_{aj} ( L ) g_{j b} ( L ) - g_{a b} ( L )
  \right], \label{BMS}
\end{equation}
with
\begin{equation}
  U_{a b j} ( L ) = \exp \left[ \tL \int_{\text{right}}  \frac{d
  \Omega_1}{4 \pi}  \left( \cW_{a b}^1 - \cW_{a j}^1 - \cW_{j b}^1 \right) \right].
  \label{Uabjdef}
\end{equation}
The boundary conditions on the BMS equation are that $g_{a b} ( 0 )
= G_{a b} ( 0 ) = 1$ for all $a$ and $b$. Importantly, this boundary condition respects any symmetry
acting on $a$ and $b$.

Before exploring the perturbative solution to the BMS equation, let us quickly consider
the symmetries of $g_{ab}(\tL)$ for different $a$ and $b$. 
The directions $a$ and $b$ can be arbitrary angles $( \theta_a, \phi_a)$ and $( \theta_b, \phi_b )$ on 2-sphere. There is an obvious
 cylindrical symmetry with respect to the hemisphere axis which makes 
 $g_{a b}(\tL)$ only depend on $\phi_b - \phi_a$.  Thus one would think there are three degrees of freedom
 in $g_{ab}(L)$. Remarkably however, the BMS equation contains a hidden $\pslr$ symmetry, and there
 is actually only one degree of freedom in $g_{ab}(L)$: the geodesic distance between $a$ and $b$
 on the \poincare disk. We explain this symmetry in Section~\ref{sec:bmsprop}.

\subsection{Perturbative check}

First, let us check that the perturbative expansion of the BMS equation for
$g_{n \nbar}$ reproduces the integrands at 2 and 3 loops that were derived in
 Sections~\ref{sec:seo} and~\ref{sec:nghmi}
by summing virtual and real corrections using the strong-energy-ordered approximation.

To work perturbatively we write, $g_{ab}(L) = \sum_{m=0}^\infty g_{ab}^{(m)}$ with $g_{ab}^{(m)}$
proportional to $L^m$, and similarly for $U_{abj}$.
Substituting $ g_{ab}^{(0)}=g_{a b}(0)=1$ and  $U_{a b j}^{(0)} = 1$, right-hand side of Eq. (\ref{BMS})
vanishes. Integrating Eq. (\ref{BMS}) we then find that there is  no $\mathcal{O} ( L )$
term in $g_{a b} ( L )$, consistent with the leading non-global
logarithm starting at 2 loops.

At order $L$, labeling the radiated gluon $2$ for convenience, we have
\begin{equation}
  U_{a b j}^{( 1 )} ( L ) = \tL \int_{\text{right}} 
  \frac{d \Omega_2}{4 \pi}  \left( \cW_{a b}^2 - \cW_{a j}^2 - \cW_{j b}^2 \right)
\label{eq:Uabj}
\end{equation}
and so
\begin{equation}
  \partial_L g_{a b}^{\left( 2 \right)} \left( L \right) = - \tL
  \int_{\text{left}}  \frac{d \Omega_1}{4 \pi}  \cW_{a b}^1 \left[
  \int_{\text{right}}  \frac{d \Omega_2}{4 \pi}  \left( \cW_{a j}^2 - \cW_{j b}^2
  - \cW_{a b}^2 \right) \right],
\end{equation}
giving
\begin{equation}
  g_{a b}^{\left( 2 \right)} \left( L \right) = - \frac{1}{2} \tL^2 \int_\Omega 1_L 2_R
  \left( \mathcal{P}_{a b}^{12} - \cW_{a b}^1 \cW_{a b}^{2 2}
  \right),
\label{eq:twoloopbms}
\end{equation}
where $\int_\Omega 1_L 2_R = \int_{\text{left}}  \frac{d \Omega_1}{4 \pi} 
  \int_{\text{right}}  \frac{d \Omega_2}{4 \pi} $.
Eq.~(\ref{eq:twoloopbms}) agrees exactly with Eq. (\ref{2loopform}).

For $g_{a b}^{\left( 3 \right)}$ we need $U_{a b j}^{\left( 2 \right)}$ and
$g_{a b}^{\left( 2 \right)}$. So,
\begin{multline}
 \partial_\tL g_{a b}^{\left( 3 \right)} \left( L \right) = \frac{1}{2} \tL^2
   \int_\Omega 1_L 2_R 3_R \cW_{a b}^1 \left( \cW_{a 1}^2 + \cW_{1 b}^2 - \cW_{a b}^2 \right)
   \left( \cW_{a 1}^3 + \cW_{1 b}^3 - \cW_{a b}^3 \right)
   \\
  - \frac{1}{2} \tL^2 \int_\Omega 1_L 2_L 3_R \cW^1_{a b}\left[ \left( \cP_{a 1}^{23} - \cW_{a 1}^2
  \cW_{a 1}^3 \right) + \left( \cP_{1 b}^{23} - \cW_{1 b}^2 \cW_{1 b}^3 \right) -
  \left( \cP_{a b}^{23} - \cW_{a b}^2 \cW_{a b}^3 \right) \right]
\end{multline}
Integrating gives
\begin{multline}
 g_{a b}^{\left( 3 \right)} \left( L \right) = \frac{1}{3!} \tL^3 \int_\Omega 1_L 2_R
   3_R \cW_{a b}^1 \left( \cW_{a 1}^2 + \cW_{1 b}^2 - \cW_{a b}^2 \right) \left( \cW_{a
   1}^3 + \cW_{1 b}^3 - \cW_{a b}^3 \right) 
   \\
  - \frac{1}{3!} \tL^3 \int_\Omega 1_L 2_L 3_R \cW^1_{a b}\left[ \left( \cP_{a 1}^{23} - \cW_{a 1}^2
  \cW_{a 1}^3 \right) + \left( \cP_{1 b}^{23} - \cW_{1 b}^2 \cW_{1 b}^3 \right) -
  \left( \cP_{a b}^{23} - \cW_{a b}^2 \cW_{a b}^3 \right) \right]
\end{multline}
which agrees with Eq. (\ref{SNG3}).

It quickly becomes clear that this is a much simpler way to generate the
integrand than following the real/virtual emission rules as in Sections~\ref{sec:seo} and
\ref{sec:nghmi}. More importantly, the BMS equation clarifies the symmetries of $g_{ab}$ which
are not at all apparent working order by order using the SEO approximation, as we will soon see.

\section{Simplifying and solving the BMS equation}
\label{sec:simp}
Before trying to iterate and integrate the BMS equation, it will be helpful to calculate
$U_{abj}$ in Eq.~\eqref{Uabjdef} exactly. This will produce a form of the measure
in the BMS equation which manifests the $\pslr$ symmetry.

\subsection{Exact solution for $U_{a b j}$}
Note that from Eq. (\ref{BMS}) the $j$ emission is always in the left
hemisphere. Thus the only relevant direction in the right hemisphere is the
hemisphere axis $n$. Thus, for the hemisphere NGLs,  we only need $U_{a n j}$ and $U_{a b j}$ with $a b$ and $j$ going left. The dipole radiator $ \cW_{a b}^j = \frac{(ab)}{(aj)(jb)}$ depends on
the round bracket from Eq. (\ref{roundbracket}) which can be expanded as
\begin{equation}
  ( ab ) = 1 - \cos
  \theta_{a b} = 1 - \cos \theta_a \cos \theta_b - \cos ( \phi_a - \phi_b
  ) \sin \theta_a \sin \theta_b \label{roundbracketex}
\end{equation}
It is helpful also to define a square bracket as the round bracket with one of the vectors reflected to the opposite hemisphere:
\begin{equation}
  [ a b ] \equiv ( \bar{a} b ) = 1 + \cos \theta_a \cos
  \theta_b - \cos ( \phi_a - \phi_b ) \sin \theta_a \sin \theta_b
\label{squaredef}
\end{equation}

Now, if $a$ and $b$ are both left, but the emission goes right, then there are no
collinear singularities in the angular integral and the dipole radiator can be easily integrated
\begin{equation}
  \int_{\text{right}}  \frac{d \Omega_1}{4 \pi} \cW_{a b}^1 = \frac{1}{2} \ln
  \frac{\left[ a b \right]}{2 \cos \theta_a \cos \theta_b} \label{Wab}
\end{equation}
Adding three of these and exponentiating with  Eq.~\eqref{Uabjdef}  leads to
\begin{equation}
  U_{a b j} ( L ) = 2^{\tL / 2} \cos^\tL \theta_j \left\{ \frac{[ a b ]}{[ a j ] [ j b ]} \right\}^{\tL / 2}
\end{equation}
Therefore, Eq.~\eqref{BMS} reduces to
\begin{multline}
  \partial_\tL g_{a b} ( L ) = \frac{1}{4 \pi} \int_0^1 d \cos  \theta_j \int_0^{2 \pi} d \phi_j 
   \frac{( a b )}{( a j) ( j b )} \\
  \times \left[ 2^{\tL / 2} \cos^L \theta_j \left\{
  \frac{[ a b ]}{[ a j ] [ j b ]} \right\}^{\tL
  / 2} g_{a j} ( L ) g_{j b} ( L ) - g_{a b} ( L) \right]
  \quad (a,b~~\text{both left}) 
  \label{BMSab}
\end{multline}
Note that when $a$ and $b$ are both left, to all orders the BMS equation only involves directions in the left
hemisphere.

When one of the directions is $n$, which is in the right hemisphere, then the integral in Eq.
(\ref{Wab}) has a collinear divergences, but $U_{a b j}$ is still finite. 
We find
\begin{equation}
  U_{a n j} \left( L \right) = 2^{\tL / 2} \cos^\tL \theta_j \left\{ \frac{\left(
  a n \right)}{[ a j ] ( j n )} \right\}^{\tL / 2},
\end{equation}
and so Eq.~\eqref{BMS} becomes
\begin{multline}
  \partial_\tL g_{a n} ( L ) = \frac{1}{4 \pi} \int_0^1 
  d \cos  \theta_j \int_0^{2 \pi} d \phi_j  \frac{ ( a n )}
  {( a j ) ( j n )} \\
  \times
  \left[ 2^{\tL / 2} \cos^\tL \theta_j \left\{
  \frac{( a n )}{[ a j ] ( j n )} \right\}^{\tL
  / 2} g_{a j} ( L ) g_{j n} ( L ) - g_{a n} ( L
  ) \right]
    \quad (a~\text{left}) 
  \label{BMSan}
\end{multline}

\pagebreak

\subsection{Symmetries of the BMS equation}
\label{sec:bmsprop}

Having evaluated $U_{abj}$ exactly, the BMS equation, as in Eq.~\eqref{BMSab} now depends only on the $g_{ab}(L)$ functions and an explicit integration measure.
In this form it is simpler to explore its symmetries.
 In the
following discussion, we will first concentrate on the BMS equation
when both $a$ and $b$ are in the left
hemisphere (as of course is the emission $j$). The case when $b$ is in the left hemisphere is similar,
but the symmetry is less obvious. We present both results in the end.

It has been observed that the BMS
equation is formally similar to the BK equation~\cite{NUPHA.B463.99,PHRVA.D60.034008},
a non-linear integro-differential equation describing gluon saturation effects. 
The BK equation enjoys a conformal symmetry
$\pslc$ in its integral measure (see, {\it e.g.}, \cite{hep-ph/0306279}), which is violated by 
initial conditions. It is therefore natural to look for a similar symmetry
in the BMS equation. 
Indeed, it has been observed that the integration
measure of the BMS equation does indeed respect $\pslc$~\cite{0903.4285,0909.0056}. 
Moreover, unlike for the BK equation, this symmetry is
not broken by the initial condition of the BMS equation. However, it is broken by 
the restriction on the integration region. As we will now explain, for the hemisphere mass case,
the restriction that radiation goes into the left hemisphere breaks the symmetry
from $\pslc$ to $\pslr$.

\begin{figure}
{
\vspace{-25mm}
  \centering
        \includegraphics[width=0.7\textwidth]{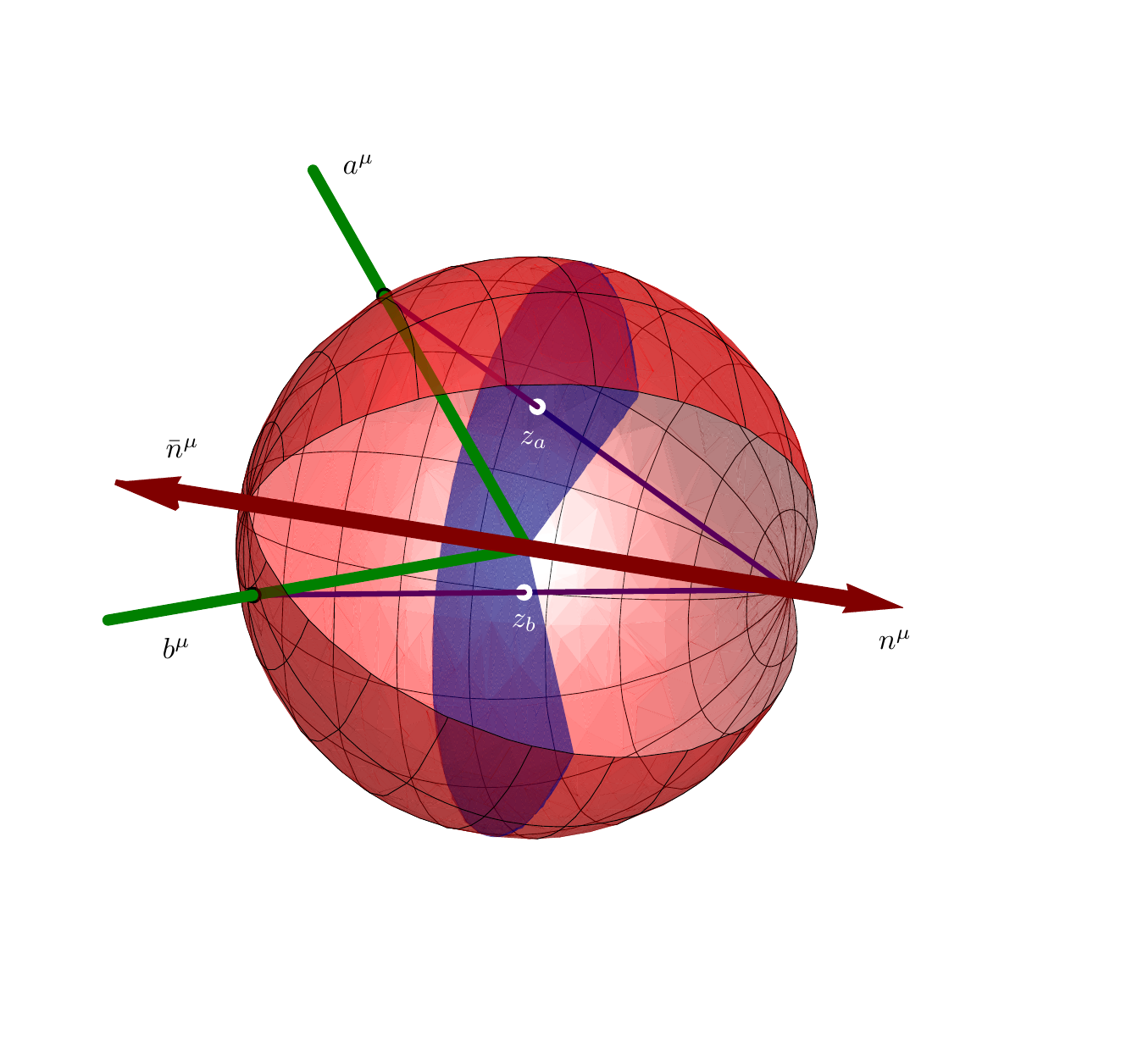}        
\vspace{-15mm}
\caption{A stereographic projection of the jet directions onto the \poincare disk reveals
the $\pslr$ symmetry of the BMS equation.}
\label{fig:proj}
}
\end{figure}

To reveal the symmetry of the BMS equation, it's useful to consider a
change of variables by stereographic projection~~\cite{0903.4285,0909.0056},
\begin{equation}
  z = \frac{\sin\theta}{1+\cos\theta} e^{i\phi}
\label{eq:z}
\end{equation}
This projection is shown in Fig.~\ref{fig:proj}.
Under the stereographic projection transformation, the full angle
space 
$(\theta,\phi)$ coordinate is mapped to the full complex plane, while
the left hemisphere, $\cos\theta>0$, is mapped to the unit disk. 

In terms of $z$, the angle from the hemisphere axis is
\begin{equation}
\cos \theta = \frac{1-|z|^2}{1+|z|^2}
\end{equation}
and the angular measure on the sphere turns into
\begin{equation}
d\Omega =  d\cos\theta d\phi = \frac{4 dz\,d\bar{z}}{(1+|z|^2)^2}
\label{eq:measure}
\end{equation}
Also, the round and square bracket inner products, in Eqs.~\eqref{roundbracketex} and \eqref{squaredef} become
\begin{align}
  (ij) =  2\frac{|z_i - z_j|^2}{( 1+|z_i|^2) (1+|z_j|^2)}, 
\qquad
  [ij] =  (ij) + 2\left(\frac{ 1-|z_i|^2|}{1+|z_i|^2}\right)\left(\frac{ 1-|z_j|^2|}{1+|z_j|^2}\right) ,
\end{align}
and the radiator times the measure becomes
\begin{equation}
d\Omega_j\, \cW_{ab}^j =d \Omega  \frac{(ab)}{(aj)(jb)} =2 dz d\bar{z} \frac{|z_a - z_b|^2}{|z_a-z_j|^2|z_j -  z_b|^2} 
\end{equation}

Recall that the angle and square brackets come from Lorentzian inner products of normalized 4-vectors on the unit sphere. Although
the sphere is Euclidean, these inner products are naturally hyperbolic. Indeed, the inner products are 
reminiscent of the hyperbolic distance measure on the \poincare disk, defined as
\begin{equation}
\met{ij}  =  \frac{|z_i- z_j|^2}{(1-|z_i|^2) (1-|z_j|^2)} = \frac{(ij)}{2 \cos \theta_i \cos \theta_j}
 \label{metdef}
\end{equation}
It then follows that 
\begin{align}
(ij ) &=  2\cos\theta_i \cos \theta_j\met{ ij}= 2\left(\frac{ 1-|z_i|^2}{1+|z_i|^2}\right)\left(\frac{ 1-|z_j|^2}{1+|z_j|^2}\right) \met{ ij}
\\
[ij] &=  2\cos\theta_i \cos \theta_j \Big(1+ \met{ ij} \Big)= 2\left(\frac{ 1-|z_i|^2}{1+|z_i|^2}\right)\left(\frac{ 1-|z_j|^2}{1+|z_j|^2}\right)
\Big(1+ \met{i j}  \Big)
\end{align}

Plugging these equations into the BMS equation for left-hemisphere NGLs,
Eq.~(\ref{BMSab}), we obtain
\begin{align}
  \partial_\tL g_{ab}(L) =& \int_{|z|<1}
  \frac{dz_j\,d\bar{z}_j}{2\pi} \frac{|z_a - z_b|^2}{|z_a-z_j|^2|z_j -
  z_b|^2}
\nn
\\
&\times \left\{  \left[\frac{1+\met{ab}}{(1+\met{aj})(1+\met{jb})}
  \right]^{\tL/2}
g_{aj}(L) g_{jb}(L) - g_{ab}(L) \right\}.
\label{eq:bmsh}
\end{align}

In this form, the symmetry of the  BMS equation under $\pslr$ is easiest to verify.
First, we note that  the radiator itself
\begin{align}
d\Omega_j\, \cW_{ab}^j =    dz_j\,d\bar{z}_j \frac{|z_a - z_b|^2}{|z_a-z_j|^2|z_j -
  z_b|^2},
\label{eq:intmea}
\end{align}
or more simply its holomorphic half,
\begin{align}
  dz_j \frac{( z_a - z_b )}{(z_a - z_j)(z_j - z_b)},
\end{align}
is invariant under (i) $z \to z + \lambda$ ($\lambda \in
\mathbb{C}$); (ii) $z \to \lambda z$, ($\lambda \neq 0$) and (iii) $z \to
- 1/z$. These symmetries generate fractional linear transformations of the form
\begin{align}
  z \to \frac{\alpha z + \beta}{\gamma z + \delta} =
  \frac{\alpha}{\gamma} + \frac{\beta - \frac{\alpha
      \delta}{\gamma}}{\gamma z + \delta},\qquad \alpha\delta
  - \beta\gamma = 1,\qquad {\rm and } \;\;\alpha,\beta,\gamma,\delta
  \in \mathbb{C}.
\label{eq:lft}
\end{align}
The matrices $\begin{pmatrix}\alpha & \beta \\ \gamma &
  \delta \end{pmatrix}$ are elements of the M\"obius group
$\pslc=\mathrm{SL}(2,\mathbb{C})/(\pm I)$, where $I$ is the
unit matrix. 
One way to understand why the radiator is invariant under M\"obius transformations, is
to recall that these transformations can be derived by
projecting the disk onto the unit sphere, rotating the sphere, and then projecting back.  

Despite the fact that the integration measure in the BMS equation respects $\pslc$, the restriction of the integration region to the left-hemisphere only ($|z|<1$ from the integration region in Eq.~(\ref{eq:bmsh})), or right-hemisphere only ($|z|>1$, see, {\it e.g.}, the integration region for $U_{abj}$ in Eq.~(\ref{eq:Uabj})), breaks $\pslc$ to $\pslr$. It is easiest to see that $\pslr$ is preserved by mapping the disk to the upper half plane, where $\pslr$
is represented by fractional linear transformations with real elements. On the disk, the subgroup of complex fractional linear transformations
preserved is spanned by matrices of the form
\begin{align}
\Gamma_d  =  \left \{ \begin{pmatrix}\alpha & \beta \\ \bar{\beta} &
  \bar{\alpha} \end{pmatrix}/(\pm I); \alpha,\beta \in \mathbb{C},
|\alpha|^2 - |\beta|^2 = 1  \right\} \label{eq:Gd}
\end{align}
These  M\"obius transformations respect the metric $\met{ij} = \frac{ (ij) }{2\cos\theta_i \cos\theta_j}$ and preserve the \poincare disk. Although they include azimuthal rotations, they are in general
not Lorentz transformations (in fact, not even  $(ab)$ is Lorentz invariant, since $a^\mu$ and $b^\mu$ have their energy component fixed to $1$ which breaks boost invariance). 

\begin{figure}[t]
  \centering 
   \includegraphics[width=0.9\textwidth]{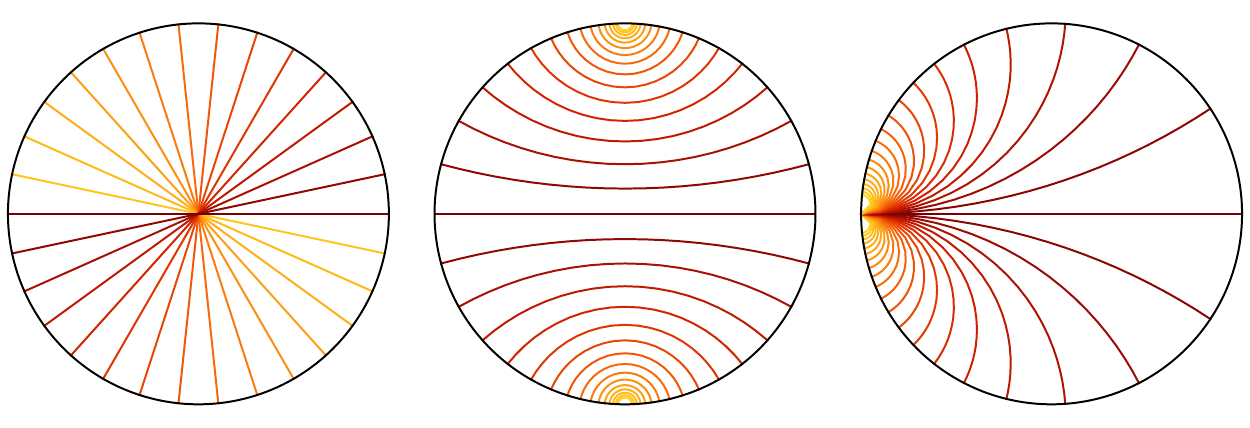}
  \caption{Elements of $\pslr$ can be visualized by their action on geodesics. Some group elements are shown.}
  \label{fig:transformations}
\end{figure}

The M\"obius  transformations are conformal mappings, preserving angles.
One way to visualize them is through their action on geodesics. Geodesics on the \poincare disk are circular arcs perpendicular to the boundary. The $\pslr$ symmetry maps geodesics to other geodesics. For example, the $x$-axis diameter is a geodesic. Transformations with $\beta=0$ and $\alpha=e^{i\phi}$ for $\phi \in \mathbb{R}$ are rotations. Transformations with $\beta\ne0$ move the origin. Some 1-parameter families of transformations are shown in Fig~\ref{fig:transformations}. To see the action of these transformations on $a$ and $b$, one can project  $a$ and $b$ to the disk, find a geodesic passing through them, transform it, then project back onto the sphere.

We  conclude that the BMS equation respects $\pslr$, and so $g_{ab}(L)$ can only depend
on the distance between $a$ and $b$ according to the metric on the \poincare disk. That is,
$g_{ab}(L)$ only depends on
\begin{equation}
g_{ab}(L) = g({\met{ab}},L)
= g\left(\frac{(ab)}{2\cos \theta_i \cos \theta_j}, L\right)
= g\left(\frac{1-\cos\theta_{ij} }{2\cos \theta_i \cos \theta_j}, L\right)
\end{equation}
 Because of this property, without loss of generality,
we can choose $z_a=0$, or $\theta_a = 0$. That is, we identify
$a=\nbar$ in the calculation. We therefore only need
\begin{equation}
  \met{\nbar j} = \frac{1-\cos\theta_j}{2\cos\theta_j},\qquad
\met{\nbar b} = \frac{1-\cos\theta_b}{2\cos\theta_b}
\label{eq:daj}
\end{equation}
This greatly simplifies the calculation of the NGLs.

\section{Perturbative calculation of NGLs to five loops}
\label{sec:ngl2}
While the symmetry of the BMS equation is clearer under stereographic
projection, we find it more convenient to perform the integrals over angles. 
It is convenient to define
\begin{equation}
  r_{ij} = \frac{1}{2}\ln \Big( 1 + \met{ij}\Big) =\frac{1}{2}\ln \frac{[ij]}{2\cos \theta_i \cos \theta_j}
 \end{equation}
 $r_{ij}$ is essentially the 1-loop Sudakov factor, Eq.~(\ref{Wab}).
 Then Eq.~\eqref{BMSab} becomes
\begin{equation}
  \partial_\tL g_{\nbar b} ( L ) = \frac{1}{4 \pi} \int_0^1 d \cos  \theta_j \int_0^{2 \pi} d \phi_j 
 \frac{(\nbar b)}{(\nbar j)(jb)} 
  \left[ e^ {\tL (r_{\nbar b} - r_{\nbar j} - r_{j b })} g_{\nbar j} ( L ) g_{j b} ( L ) - g_{\nbar b} ( L) \right]
  \label{BMSab2}
\end{equation}
To obtain the $m$-loop NGLs, we
expand Eq.~(\ref{BMSab2}) recursively.
Recalling that $g_{ab}^{(0)} = 1$ and $g_{ab}^{(1)} = 0$, we get
\begin{align}
      \partial_\tL g^{(2)}_{\nbar b}(L) =& \frac{1}{4\pi} \int^1_0 d\cos\theta_j
  \int^{2\pi}_0 d\phi_j 
   \frac{(\nbar b)}{(\nbar j)(jb)} 
 (r_{\nbar b} - r_{\nbar j} - r_{j b }),
 \\
      \partial_\tL g^{(3)}_{\nbar b}(L) =& \frac{1}{4\pi} \int^1_0 d\cos\theta_j
  \int^{2\pi}_0 d\phi_j 
     \frac{(\nbar b)}{(\nbar j)(jb)} 
 \left[ \frac{\tL^2}{2} (r_{\nbar b} - r_{\nbar j} - r_{j b })^2
 + g^{(2)}_{\nbar j } +   g^{(2)}_{jb}  -  g^{(2)}_{\nbar b}  
 \right],
\\
      \partial_\tL g^{(4)}_{\nbar b}(L) =& \frac{1}{4\pi} \int^1_0
      d\cos\theta_j 
  \int^{2\pi}_0 d\phi_j 
     \frac{(\nbar b)}{(\nbar j)(jb)} 
\nn
\\
&\times
\left[
 \frac{\tL^3}{6}(r_{\nbar b} - r_{\nbar j} - r_{j b })^3
 +
 \tL ( r_{\nbar b} - r_{\nbar j} - r_{j b })
 (g^{(2)}_{ \nbar j} + g^{(2)}_{jb}) 
 + g^{(3)}_{\nbar j } +   g^{(3)}_{jb}  -  g^{(3)}_{\nbar b}  
 \right]
\end{align}
We note that at each order, there are exactly
one azimuthal angle integral and one polar angle integral to be done,
once the lower order NGLs are known. Also these integrals are finite,
although there are singular denominators at each order. 

\subsection{Azimuthal integrals}
The azimuthal integrals can be performed using contour integration.
 As a simple yet non-trivial example, consider an azimuthal
integral required for the 2-loop NGLs,
\begin{equation}
\az_2 = \int^{2\pi}_0 \frac{d\phi_j}{2\pi} \frac{1}{(jb)} \ln \Big(1 + \met{jb}\Big)
\label{eq:aex}
\end{equation}
After the standard change of variables, $ t = e^{i \phi_j}$, this becomes
%
\begin{equation}
\az_2=  - \frac{2}{\sin\theta_j\sin\theta_b} \oint_C \frac{dt}{2\pi i}
  \frac{1}{ (t - t_+ )(t - t_-)} \ln \frac{1 + \cos\theta_j
    \cos\theta_b - \sin\theta_j \sin\theta_b \left(\frac{1}{2t} +
      \frac{t}{2} \right) }{2\cos\theta_j \cos\theta_b},
\end{equation}
where the integral contour is the unit circle, and
\begin{equation}
  t_+ = \frac{1 - \cos\theta_j \cos\theta_b + |\cos\theta_j -
    \cos\theta_b|}{\sin\theta_j \sin\theta_b},\qquad 
  t_- = \frac{1 - \cos\theta_j \cos\theta_b - |\cos\theta_j -
    \cos\theta_b|}{\sin\theta_j \sin\theta_b}
\end{equation}

The factor $\frac{1}{(t-t_+)(t-t_-)}$ contains two single poles, but only the pole at
$t_-$ is within the unit circle. Furthermore, since $t_+$ and $t_-$
are the solutions of the equation $\met{jb}=0$, or more precisely
$(jb)=0$, it follows that the logarithmic factor $\ln(1+\met{jb})$
vanishes at $t_+$ and $t_-$.

\begin{figure}[t]
  \centering
  \includegraphics[]{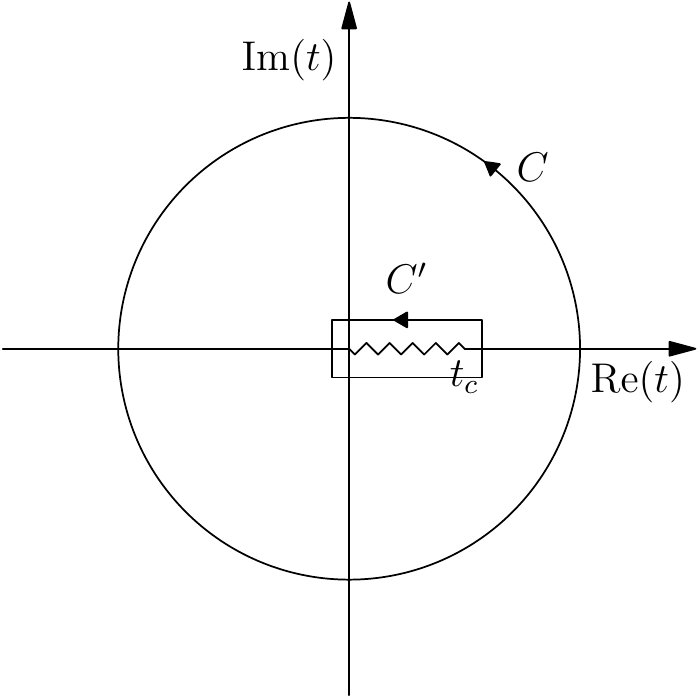}
  \caption{Integral contour for the $t$ integral}
  \label{fig:contour}
\end{figure}

The logarithm function contains a branch
cut on the negative real axis. Solving for the inequality $\met{jb}<-1$,
we find that the branch cut in the $t$-complex plane is from $0$ to
$t_c = \frac{1- \cos\theta_j - \cos\theta_b +
  \cos\theta_j\cos\theta_b}{\sin\theta_j\sin\theta_b}$. Since the
integrand is analytic elsewhere within the unit circle, we can shrink
the contour $C$ to $C'$ as shown in Fig.~\ref{fig:contour}, without changing the value of the
integral. The original azimuthal angle integral is therefore traded
for a
line integral, 
\begin{equation}
\az_2 = 
\frac{4\pi i}{\sin\theta_j\sin\theta_b} \int^{t_c}_0 \frac{dt}{2\pi i}
\frac{1}{ (t-t_+)(t-t_-)},
\end{equation}
where we have made use of the fact that the discontinuity of log
function on the negative real axis is $2\pi i$. This line integral can
then be trivially done, giving a simple result
\begin{equation}
  \int^{2\pi}_0 \frac{d\phi_j}{2\pi} \frac{1}{(jb)} \ln (1 + \met{jb}) =
  \frac{\Big(1+2\met{b\nbar}\Big)\Big(1+2\met{j\nbar}\Big)}{2 \Big(\met{b\nbar} - \met{j\nbar}\Big)} \ln \left(\frac{
      1+ \met{b\nbar}}{1+\met{j\nbar}} \right).
\label{eq:azi1}
\end{equation}
We have introduced $\nbar$ into this solution using Eq.~\eqref{eq:daj} to manifest the $\pslr$ 
invariance.

The azimuthal integral required for the  3-loop NGL can be done using the same method. The result is
\begin{multline}
 \int^{2\pi}_0 \frac{d\phi_j}{2\pi} \frac{\ln \met{jb} \ln(1+\met{jb}) +
    \mathrm{Li}_2 (-\met{jb})}{(jb)}  = \frac{1}{\cos\theta_b -
    \cos\theta_j} \\
    \times \left[ \ln\frac{1+\met{j\nbar}}{1+\met{b\nbar}} \ln
    \frac{ \met{j\nbar}+ \met{b\nbar} + |\met{j\nbar} - \met{b\nbar}
      |}{2} + \mathrm{Li}_2(-\met{j\nbar}) - \mathrm{Li}_2 (-\met{b\nbar})
\right].\label{eq:azi3T}
\end{multline}
The 4-loop azimuthal integral is in Appendix \ref{app:azi}.

For all the azimuthal integrals we consider, the integrand is of uniform transcendentality.
The azimuthal integrals are all nonsingular and do not  change the transcendentality.

\subsection{Polar integrals, GPLs, symbols and coproducts}
Once all the azimuthal angle integrals are done, which is straightforward with contour integration,
all that remains are the polar angle integrals. It turns
out to be useful to make another change of variables using
Eq.~(\ref{eq:daj}),
\begin{equation}
  \int^1_0 d\cos\theta_j \frac{(b\nbar)}{(j\nbar)} =
 \int^\infty_0 d \met{j\nbar} \frac{2\met{b\nbar}}{\Big(1+2\met{b\nbar}\Big)\Big(1+2\met{j\nbar}\Big)\met{j\nbar}}.
    \label{eq:cos1}
\end{equation}
 At 2 and 3 loops, the $\met{j\nbar}$ integral can be done straightforwardly using, {\it
  e.g.}, \texttt{Mathematica}.
To go beyond that, we note that after using partial fractions, the polar integrand has the form
of an iterated integral:
\begin{equation}
  \int^1_0 d\cos\theta_j \frac{(b\nbar)}{(j\nbar)}
  \frac{1}{\cos\theta_b - \cos\theta_j} =
   \int^\infty_0 d\met{j\nbar} \left(  \frac{1}{ \met{j\nbar}} -
    \frac{1}{\met{j\nbar}- \met{b\nbar}} \right) 
    \label{eq:cos2}
\end{equation}
This is not surprising, since we are solving the BMS equation by iteration. The
iterated form nevertheless suggests that we might be able to exploit recent developments in techniques using coproducts and Goncharov polylogarithms (GPLs)~\cite{math/0103059,1105.2076} to compute them. We now briefly review some of the relevant mathematics.

Recall that the classical polylogarithms are defined iteratively by
\begin{equation}
\Li_k(x) = \int_0^x \frac{d  t}{t} \, \Li_{k-1} ( t)
\label{Lidef}
\end{equation}
with $\Li_1(x) = -\ln(1-x)$. The GPLs are defined as a generalization of this
\begin{equation}
  \label{eq:Gdef}
  G(w_1,\dots,w_n;x) = \int^x_0 \frac{dt}{t-w_1} G(w_2,\dots ,w_n;t),
\end{equation}
with $G(;x)=1$. The set of $n$ complex numbers $\{w_1,\dots,w_n\}$ is
called the index vector of the GPL, where at least one entry is
nonzero. In the case that all the entries are zeros, the GPL is
defined to be
the $n^{\text{th}}$ power of $\ln x$, where $n$ is the length of the index
vector, also called the weight of the GPL,
\begin{equation}
  G(\underbrace{0,\dots,0}_{n};x) = \frac{1}{n!}\ln^nx
\label{eq:glog}
\end{equation}
Classical polylogarithms consist of a subset of the GPLs,
\begin{equation}
  G(\underbrace{a,\cdots,a}_{n};x) = \frac{1}{n!} \ln^n\left(1 -
    \frac{x}{a} \right),
\qquad
G(\underbrace{0,\cdots,0}_{n-1}, a; x) = - \mathrm{Li}_n
\left(\frac{x}{a} \right)
\label{eq:gli}
\end{equation}

If a given integrand can be
written so that the integration variable shows up in the
argument of a  GPL and not in its index vector, 
then the result for the integral can simply be read off using Eq.~(\ref{eq:Gdef}).
In our case,  after the azimuthal integrals are done,  the integrands
are, in general,  complicated
combinations of classical polylogarithms. These  classical
polylogarithms can be converted into GPLs using Eq.~(\ref{eq:glog}) and
(\ref{eq:gli}). 
However, the resulting GPLs representation of the
integrand will not be in the form of Eq.~(\ref{eq:Gdef}). Instead, the
integration variable shows up both in the argument and in the index vector
in a complicated manner. It is therefore necessary to use functional identities obeyed by the GPLs to 
massage the
integrand into the canonical form, Eq.~(\ref{eq:Gdef}).

 A very useful
tool for simplifying the integrand is the technique of
symbols~\cite{math/0208144}, first introduced in physics in the simplification of 2-loop
6-particle remainder function in ${\mathcal N}=4$ Super Yang-Mills
theory~\cite{1006.5703}. The idea is to map the complicated combination of
GPLs to a tensor algebra over the group of rational
functions, by computing its symbol. In this way, functional identities obeyed by the GPLs are
mapped to simpler algebraic identities. We then simplify the symbol
using these algebraic relation and finally reconstruct the original expression
in the desired form using its symbol.

The symbol acts naturally on iterated integrals of the form
\begin{equation}
T_k = \int_a^b d \ln R_1 \circ \cdots \circ d \ln R_k 
\end{equation}
where $R_i(t)$ are rational functions. The iteration is defined recursively as
\begin{equation}
 \int_a^b d \ln R_1 \circ \cdots \circ d\ln R_k 
= \int_a^b  d \ln R_n (t) \Big(\int_0^t d \ln R_1 \circ \cdots \circ d \ln R_{n-1} \Big)
\end{equation}
Both classical polylogarithms and GPLs, defined by Eqs.~\eqref{Lidef} and \eqref{eq:gli}, are given by iterated integrals in this category, with $R_k(t) = 1-t$ and $R_k(t) = t-w_k$ respectively.  
The symbol of
an iterated integral is denoted as
\begin{equation}
\cS[ T_k ] =R_1 \otimes \cdots \otimes R_k
\end{equation}
so that
\begin{equation}
\cS[\Li_k(x) ] = -(1-x) \otimes \underbrace{x \otimes \cdots \otimes x}_{k-1}
\end{equation}
and
\begin{align}
  \mathcal{S}\left[ G\left( a_1,\dots,a_n;x \right) \right] = \left( 1
    - \frac{x}{a_n} \right) \otimes \dots \otimes \left( 1-
    \frac{x}{a_1} \right)
\label{eq:Gsym}
\end{align}
Another important property of the symbol is that 
\begin{equation}
R_1 \cdots \otimes (R_a R_b) \otimes R_k =
 R_1 \cdots \otimes R_a\otimes \cdots R_k
+
 R_1 \cdots \otimes R_b\otimes \cdots R_k
\end{equation}
and $\cS[ c ] = 0$ for constants $c$.

As an example, we consider the polar angle integral
over the azimuthal-averaged integrand at 3 loops. Specifically, we are interested in the following
integral
\begin{equation}
\frac{1}{\tL^3} g^{(3)}_{\nbar b} (L)= - \frac{1}{24}\int^\infty_0
d\met{j\nbar}  \left(\frac{1}{\met{j\nbar}} - \frac{1}{\met{j\nbar} -
    \met{b\nbar} } \right)
 \Phi_3 (\met{b\nbar},\met{j\nbar})
\label{eq:i3}
 \end{equation} 
 $ \Phi_3 (\met{b\nbar},\met{j\nbar})$ is a piecewise smooth
function of $\met{j\nbar}$ for fixed $\met{b\nbar}$. Let
\begin{align}
  u_1 = \met{b\nbar}, \qquad u_2 = \met{j\nbar},
\end{align}
we then have
\begin{equation}
\left. \Phi_3 (u_1, u_2) \right|_{u_1 > u_2 }
= \ln (1 + u_2) \left(\ln \frac{u_1(1+u_2)}{u_2 (1+u_1)} \right),
\label{eq:phi3a}
\end{equation}
and
\begin{multline}
\left. \Phi_3 (u_1, u_2) \right|_{u_1 < u_2 }
= \ln u_1 \ln(1+u_1) - 2\ln^2 (1+u_1) + \ln (1+u_1) \ln u_2 + 3
\ln(1+u_1) \ln (1+u_2)
\\
 -  2 \ln u_2 \ln (1+u_2) - \ln^2 (1+u_2) + 2
\mathrm{Li}_2 (-u_1) - 2 \mathrm{Li}_2 (-u_2) 
\label{eq:phi3b}
\end{multline}
The integral in Eq.~(\ref{eq:i3}) is naturally split into two
pieces,
\begin{align}
  \cI_3 = - \frac{1}{24} \int^{u_1}_0 d u_2 \left( \frac{1}{u_2} -
    \frac{1}{u_2 - u_1} \right) \Phi_3 (u_1,u_2)
\label{eq:ii3}
\end{align}
and
\begin{align}
  \mathcal{J}_3 = - \frac{1}{24} \int^{\infty}_{u_1} d u_2 \left( \frac{1}{u_2} -
    \frac{1}{u_2 - u_1} \right) \Phi_3 (u_1,u_2)
\label{eq:jj3}
\end{align}
For simplicity, we only show details for the computation of 
$\cI_3$. $\mathcal{J}_3$ can be obtained in almost the same way, after changing  variables to move the lower bound of the integration range to $0$.

To proceed,
we first compute the symbol of $\left. \Phi_3(u_1,u_2) \right|_{u_1>u_2}$,
\begin{multline}
\cS \left[ \left.\Phi_3(u_1,u_2) \right|_{u_1>u_2} 
\right]
=
u_1 \otimes (1+u_2) - (1+u_1) \otimes (1+u_2) - u_2 \otimes (1+u_2) +
(1+u_2) \otimes u_1
\\
 - (1+u_2) \otimes (1+u_1) - (1+ u_2) \otimes u_2
+
2 \left[  (1+u_2) \otimes (1+u_2)\right]
\label{eq:phi3lab}
\end{multline}
It is straightforward to find a set of GPLs with the same symbol.
 A simple
algorithmic approach is given in Ref.~\cite{1302.4379}. The important
observation is that the symbol of a GPL with argument $x$ and an $x$-independent index
vector consists of a single
term, as in Eq.~(\ref{eq:Gsym}). Note that $x$ shows up
in every entry of the symbol in Eq.~(\ref{eq:Gsym}). To match the symbol of
Eq.~(\ref{eq:phi3lab}), we start from the terms where the next
integral variable, $u_2$, shows up in every entry of the
symbol. For example, the following GPL has exactly the same symbol
as the last term in Eq.~(\ref{eq:phi3lab})
\begin{align}
  \mathcal{S}[G(-1,-1;u_2)] = (1+u_2)  \otimes  (1+u_2)
\end{align}
We then proceed to reconstruct the symbol where at least one entry is
independent of $u_2$, {\it e.g.}, $(1+u_2) \otimes
(1+u_1)$. From Eq.~(\ref{eq:Gsym}) we know that such a symbol cannot
correspond to a single GPL where $u_2$ only shows up in the argument
but not in the index vector. They can, however, arise from the product of
two GPLs, $G(-1; u_2) G(-1; u_1)$. In fact, the
product of GPLs gives not a single term but two terms, which match
exactly with part of the symbol in Eq.~(\ref{eq:phi3lab}),
\begin{align}
  \mathcal{S}[G(-1;u_2) G(-1; u_1)] = (1 + u_2) \otimes (1+u_1) + (1+
  u_1) \otimes (1+u_2)
\end{align}
Such procedure can be iterated until  the entire symbol has been
reconstructed. The result is the following ansatz,
\begin{multline}
  \Phi^G_3 (u_1,u_2) = G(0; u_1)G(-1; u_2) - G(-1; u_1) G(-1; u_2) \\
-
  G(-1,0; u_2)
 - G(0,-1; u_2)
 + 2G(-1,-1, u_2)
\end{multline}
However, since the symbol maps all
constants to zero, we cannot yet  conclude that $\Phi_3
\left. (u_1,u_2) \right|_{u_1 > u_2
} = \Phi^G_3 (u_1,u_2)$. The two functions have uniform transcendentality, but they  may differ
by  terms of transcendentality $2$, such as $\pi^2$, or terms like $i \pi \times \ln$.
 In our current case, 
both $\Phi_3$ and $\Phi^G_3$ are real for $u_1> u_2>0$, so they can not differ by terms like $i\pi
\times\ln$. Instead, they could differ by a term proportional to
$\pi^2$,
\begin{align}
\left.  \Phi_3 (u_1,u_2) \right|_{u_1>u_2}= \Phi^G_3 (u_1,u_2) + c\, \pi^2.
\end{align}
The rational number $c$ can be easily fixed by computing the two sides of the
above equation numerically for some $u_1$ and
$u_2$\footnote{Efficient numerical evaluation of GPLs can be
  done by \texttt{GiNaC}~\cite{cs/0004015}.}. It turns out that
$c=0$. We have thus fully
reconstructed $\Phi_3(u_1,u_2)$ for $u_1>u_2$ into the canonical
form, including the constant term.

Now the integral in Eq.~(\ref{eq:ii3}) can be done almost trivially,
using the iterative definition of GPLs, Eq.~(\ref{eq:Gdef}).
For example, 
\begin{align}
\int^{u_1}_0 du_2 \frac{1}{u_2 - u_1} G(-1,0; u_2) = G(u_1, -1,0; u_1)
\end{align}
Here one needs to be careful because the resulting GPL, 
$ G(u_1,-1,0; u_1)$, is logarithmically divergent. In
general, when the first entry of a GPL coincides with its argument,
there is a logarithmic divergence in it, as evident from the
iterational definition, Eq.~(\ref{eq:Gdef}). However, since the
original integral Eq.~(\ref{eq:i3}) is finite, as can be checked
numerically, such logarithmic divergences must be spurious and must
cancel against similar logarithmic divergences from other terms. A
simple method~\cite{1203.0454} to deal with such spurious logarithmic divergence is to
isolate them using shuffle identities of GPLs~\cite{shuffle}:
\begin{align}
  G(a_1, \dots, a_{n_1}; x) G(a_{n_1+1},\dots, a_{n_1+n_2}; x) =
  \sum_{\sigma \in \Sigma(n_1,n_2) } G(a_{\sigma(1)}, \dots,
    \sigma_{n_1+n_2}; x),
\end{align}
where the summation is over all different permutations in which the
relative ordering of the sets $\{a_1,\dots, a_{n_1}\}$ and
$\{a_{n_1+1}, \dots, a_{n_1+n_2} \}$ are preserved. Applying these
shuffle identities to $G(u_2,-1,0; u_2)$, we obtain
\begin{align}
  G(u_1,-1,0; u_1) = G(u_1; u_1) G(-1,0; u_1)  - G( -1, u_1, 0; u_1) - 
G( -1,0, u_1; u_1)
\end{align}
Now the logarithmic divergent term $G(u_1; u_1)
 G(-1,0; u_1)$ is isolated. The final result for $\cI_3$ is then given
 by
 \begin{multline}
\cI_3 = 
\frac{1}{24} G\left(-1;u_1\right) G\left(-1,u_1;u_1\right)-\frac{1}{24} G\left(0;u_1\right)
   G\left(-1,u_1;u_1\right)+\frac{1}{24} G\left(-1;u_1\right)
   G\left(0,-1;u_1\right)
\\
-\frac{1}{24}
   G\left(0;u_1\right) G\left(0,-1;u_1\right)
-\frac{1}{12} G\left(-1,-1,u_1;u_1\right)+\frac{1}{24}
   G\left(-1,0,u_1;u_1\right)
\\
-\frac{1}{12} G\left(-1,u_1,-1;u_1\right)+\frac{1}{24}
   G\left(-1,u_1,0;u_1\right)
-\frac{1}{12} G\left(0,-1,-1;u_1\right)+\frac{1}{24}
   G\left(0,-1,0;u_1\right)
\\
+\frac{1}{24} G\left(0,-1,u_1;u_1\right)+\frac{1}{24}
   G\left(0,0,-1;u_1\right)+\frac{1}{24} G\left(0,u_1,-1;u_1\right)
 \end{multline}

Doing the integral for $\mathcal{J}_3$ in the same way, we obtain the
final result for Eq.~(\ref{eq:i3}),
\begin{multline}
\frac{1}{\tL^3} g^{(3)}_{\nbar b}(L) =  \frac{\pi^2}{36} G(-1;u_1) -
\frac{1}{4} G(-1,-1,-1;u_1) \\
+ \frac{1}{4} G(-1,-1,0;u_1) 
+ \frac{1}{12}G(-1,0,-1;u_1)
 - \frac{1}{12}G(-1,0,0;u_1)
\end{multline}
where  $u_1 = \met{\nbar b}$.
In terms of classical polylogarithms this is
\begin{multline}
g^{(3)}_{\nbar b}(L) 
= \tL^3 \Bigg[ \frac{\pi^2}{72} \ln(1+u_1) - \frac{1}{24} \ln^2 u_1\ln(1+u_1) +
\frac{1}{12} \ln u_1 \ln^2 (1+u_1) - \frac{1}{36} \ln^3 (1+u_1)
\\
- \frac{1}{12} \ln u_1 \mathrm{Li}_2(-u_1) + \frac{1}{12} \ln(1+u_1)
\mathrm{Li}_2 (-u_1) + \frac{1}{12} \mathrm{Li}_3(-u_1) - \frac{1}{12}
\mathrm{Li}_3 \left(\frac{1}{1+u_1} \right) + \frac{\zeta(3)}{12}\Bigg],
\end{multline}
This result agrees with what we find by direct integration of the
3-loop integrand using  \texttt{Mathematica}.

At 4 loops, the polar integral cannot be done directly, and we find the use of symbols
to be necessary. 
One additional complication beyond 3 loops is that symbols does not
fix the functional form of the original function ({\it e.g.}, there
can be terms like $\zeta(2) \log$, which is mapped to zero under the symbol).  Fortunately, these terms can be obtained using
a generalization of the symbol called the
coproduct~\cite{math/0208144,1102.1310}, whose application in the
context of scattering amplitudes is nicely demonstrated in
Ref.~\cite{1203.0454}. We provide an example in Appendix \ref{app:coproduct} which illustrates
the use of the coproduct in our calculation. The result for $g_{ab}^{(4)}(L)$ is given
in Appendix~\ref{app:gab}.

\subsection{Analytical results for NGLs at fixed order}
\label{sec:fongl}
The formulas for $g_{ab}(L)$ with $a$ and $b$ in the left hemisphere
at up to 4 loops are given in  Appendix~\ref{app:gab}. When $b$ is in the right hemisphere, aligned with the hemisphere axis $n$,
the formulas are simpler. Defining  $y = \met{a\nbar}=\frac{1-\cos\theta_a}{2\cos\theta_a}$, we find
\begin{align}
  \frac{1}{\tL^2} g^{(2)}_{an}(L)= & - \frac{\pi^2}{24},
\label{nglLR2}
\\
\frac{1}{\tL^3} g^{(3)}_{an}(L) = &  \frac{\zeta(3)}{12},
\label{nglLR3}
\\
\frac{1}{\tL^4}g^{(4)}_{an}(L) = & \frac{\pi^4}{34560} -
\frac{\pi^2}{576} G(0,-1;y) - \frac{1}{96} G(0,-1,-1,-1;y) +
\frac{1}{96} G(0,-1,0,-1;y)
\nn
\\
=& \frac{\pi^4}{6912} - \frac{1}{576} \ln(-y) \ln^3(1+y) +
\frac{\pi^2}{576} \mathrm{Li}_2 (-y) + \frac{1}{192}
\mathrm{Li}_2(-y)^2 
\nn
\\
&
- \frac{1}{192} \ln^2 (1+y) \mathrm{Li}_2(1+y) + \frac{1}{96} \ln(1+y)
\mathrm{Li}_3 (1+y)  - \frac{1}{96} \mathrm{Li}_4(1+y)
\nn
\\
& - \frac{1}{48} S_{2,2}(-y),
\label{nglLR4}
\end{align}
where the functions $S_{2,2}(-y)$ is the Nielsen
polylogarithm. 

It is perhaps worth making a few comments about these results and their calculation:
\begin{itemize}
\item The perturbative expansion of the NGLs have uniform degree of
  transcendental weight at each order\footnote{The
    transcendental weight is $1$ for $\pi$, and $n$ for $\zeta(n)$.}. At $n$ loops, the transcendentality weight is $n$.
\item At 2 and 3 loops, the opposite hemisphere NGLs $g_{an}$ is
independent of the dipole directions. There is a low-order accident as there
is dependence on $\met{an}$ at 4 loops and beyond.

\item The asymptotic behavior of the  NGLs is straightforward to extract.  In the limit $x=\met{ab} \to 0$, $a$ and
$b$ coincide. In that limit, we find
  \begin{equation}
    \lim_{x\to 0} g^{(n)}_{ab}(L) = 0 + \mathcal{O}(x), \qquad n=2,3,4.
  \end{equation}
In the limit of $x\to \infty$, either $a$ or $b$ becomes perpendicular
to the hemisphere axis. The asymptotic behavior of the same hemisphere
NGLs in that limit is given by
\begin{align}
  \lim_{x\to \infty} g^{(2)}_{ab}(L) =& -\frac{\pi^2}{24} \tL^2 +
  \mathcal{O}\left( \frac{1}{x} \right),
\\
  \lim_{x\to \infty} g^{(3)}_{ab}(L) =& \frac{\zeta(3)}{12} \tL^3 +
  \mathcal{O}\left( \frac{1}{x} \right),
\\
  \lim_{x\to \infty} g^{(4)}_{ab}(L) =& \left(-\frac{\pi^4}{5760}  -
  \frac{\zeta(3)}{48} \ln x\right) \tL^4 +
  \mathcal{O}\left( \frac{1}{x} \right),
\end{align}
For the opposite hemisphere NGLs, the $y\to 0$ limit gives exactly the
hemisphere NGLs, because $a$ coincides with $\nbar$ in that limit. The $y\to \infty$ limit is given by
\begin{align}
  \lim_{y\to \infty} g^{(2)}_{an}(L) = &- \frac{\pi^2}{24} \tL^2 +
  \mathcal{O}\left( \frac{1}{y} \right),
\\
  \lim_{y\to \infty} g^{(3)}_{an}(L) =& \frac{\zeta(3)}{12} \tL^3 +
  \mathcal{O}\left( \frac{1}{y} \right),
\\
  \lim_{y\to \infty} g^{(4)}_{an}(L) =& \left(\frac{\pi^4}{5760}  -
  \frac{\zeta(3)}{48} \ln y\right) \tL^4 +
  \mathcal{O}\left( \frac{1}{y} \right),
\end{align}
Intriguingly, the opposite hemisphere and same hemisphere NGLs
exhibit the same logarithmic divergence for large $x$ or $y$, with the same
slope and opposite intercept.
\end{itemize}

\begin{figure}[t]
\begin{center}
\hspace{-5mm}
  \includegraphics[width=0.45\textwidth]{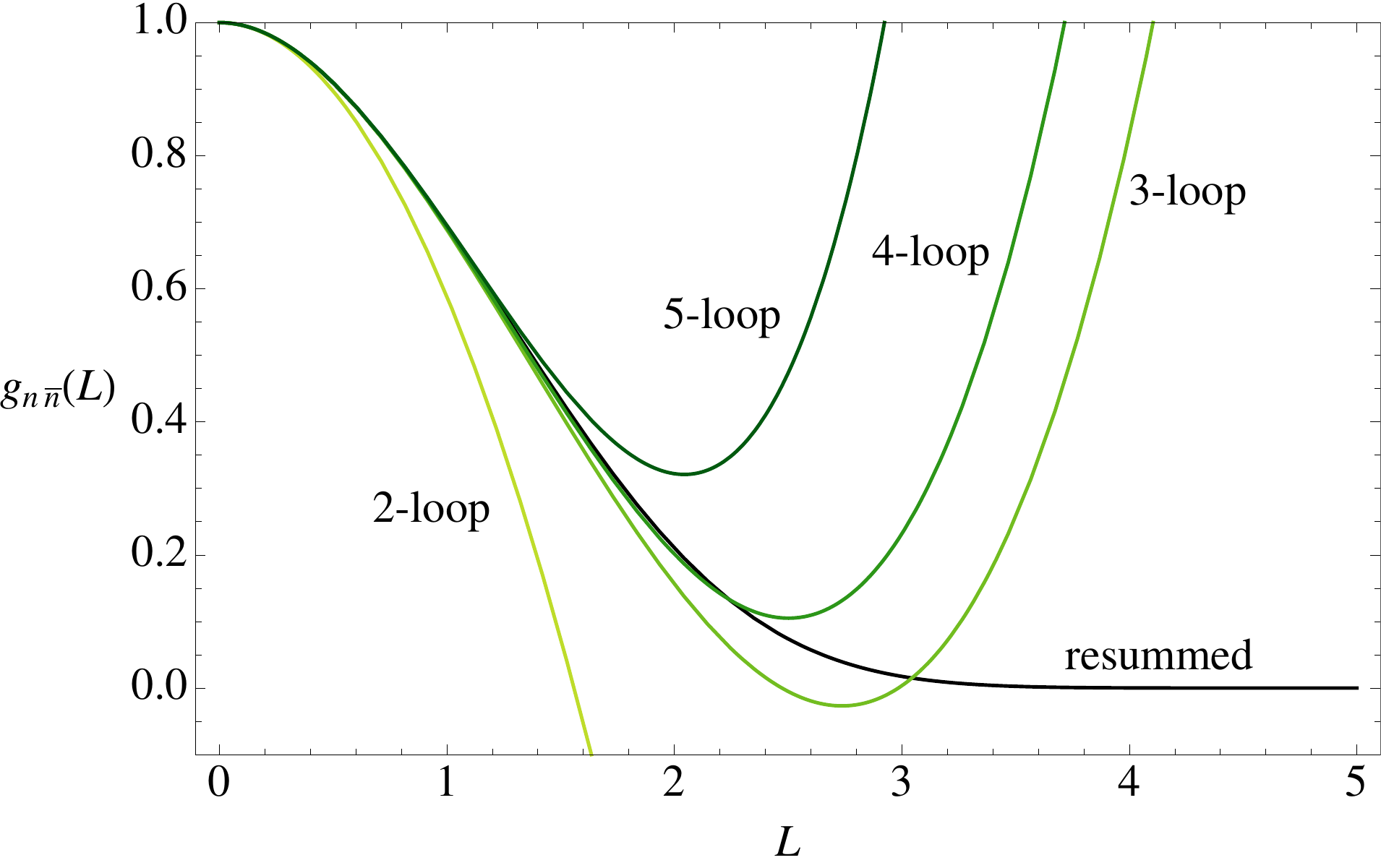}
  \hspace{5mm}
  \includegraphics[width=0.45\textwidth]{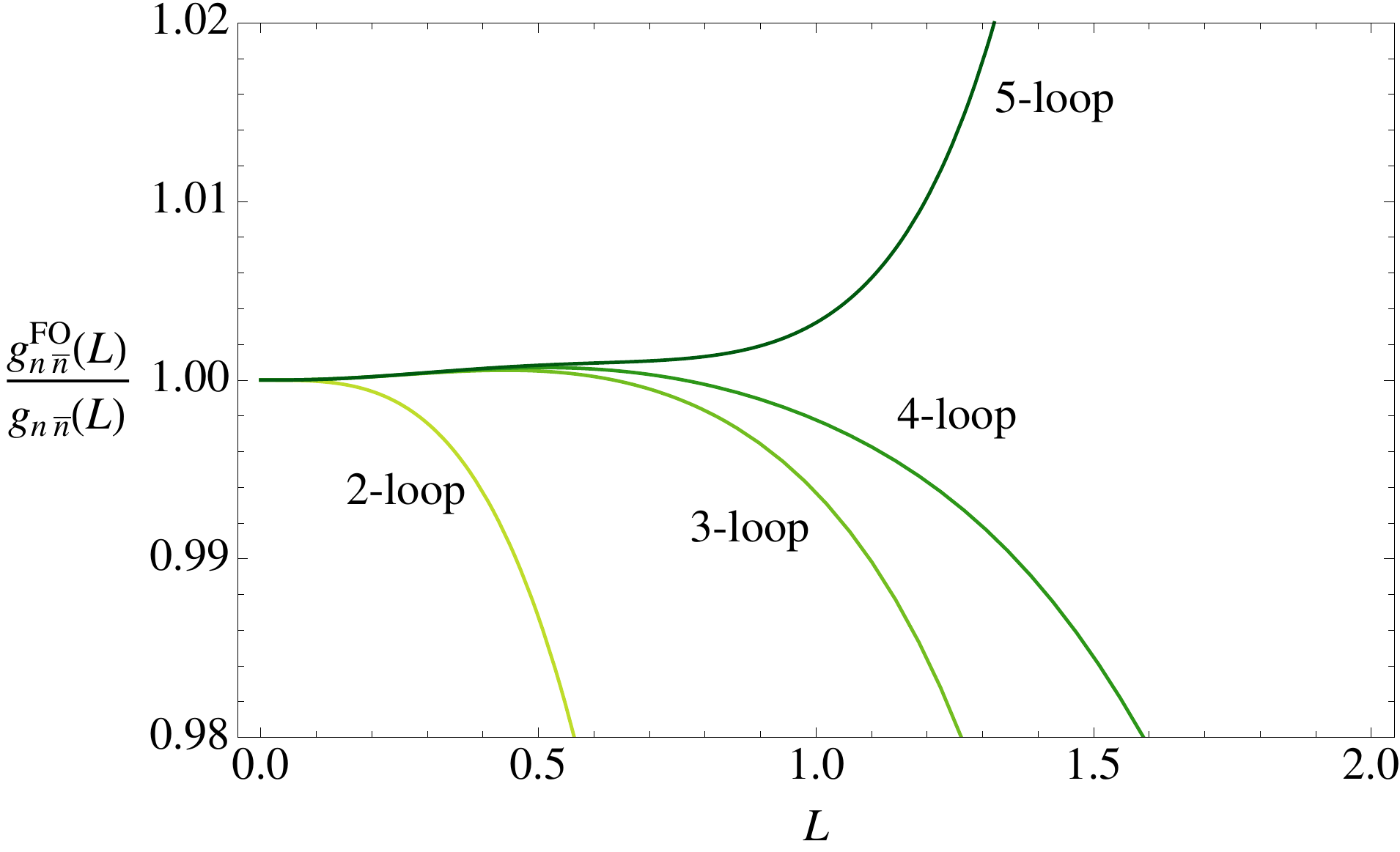}
  \caption{ Comparisons of the complete, resummed, leading NGL series
    for the hemisphere mass distribution, $g_{n\nbar}(L)$, to its
    fixed-order approximation at up to 5 loops. The resummed distribution is computed by numerically solving the BMS equation. The fixed order analytical expansions are given in Eq.~\eqref{fiveloop}.
 On the left, the numerical solution is labelled ``resummed''. The
 right plot shows the fixed order approximations relative to this
 resummed result in the region $0<\tL<2$.
  \label{fig:conv} 
  }
\end{center}
\end{figure}

Finally, using the analytic results for the opposite hemisphere and same
hemisphere NGLs up to and including 4 loops, we can calculate the
hemisphere NGLs through 5 loops. The result is
\begin{equation}
\boxed{
  g_{n\nbar}(L) =1 -\frac{\pi^2}{24}\tL^2 + \frac{\zeta(3)}{12} \tL^3 +
  \frac{\pi^4}{34560} \tL^4 + \left( -\frac{\pi^2 \zeta(3)}{360} +
  \frac{17\zeta(5)}{480} \right)\tL^5 + \dots
  }
  \label{fiveloop}
  \end{equation}
Numerically, it can be written as
\begin{equation}
  g_{n\nbar}(L) = 1-0.411233512 \tL^2 + 0.10017141 \tL^3 + 0.0028185501
  \tL^4 + 0.0037694522 \tL^5 + \dots
  \label{coeffs}
\end{equation}
Note that the 5-loop coefficient is actually larger than the as the 4-loop coefficient. Perhaps this is because the 4-loop coefficient is unusually small. In any case, it suggests that the series may not be convergent beyond $\tL=1$. Plots of the approximations of $g_{n\nbar}(L)$ at up to 5 loops and a comparison to the exact (that is, numerically resummed) result are shown in Fig.~\ref{fig:conv}.
We discuss the calculation of the resummed result in the next section.

\section{Resummation}
\label{sec:resum}
An exact solution to the  hemisphere BMS equation, Eq.~\eqref{BMS},
would resum the leading hemisphere NGL. While we cannot solve this equation analytically, finding
a numerical solution is straightforward. Before discussing the numerically approach, we explore
an iterative approach to the resummed solution, finding an exact solution in the first nontrivial
case.

\subsection{Two loop resummation}
Rather than expanding the BMS equation to fixed order and integrating, we can iterate
the equation in an alternative manner. Following~\cite{Banfi:2002hw},   we first
rewrite Eq.~\eqref{BMS} as
\begin{equation}
  \partial_\tL g_{a b} ( L ) = g_{a b} ( L )
  \int_{\text{left}}  \frac{d \Omega_j}{4 \pi} W_{a b}^j \left[ U_{a b j}
  ( L ) - 1 \right] + \int_{\text{left}}  \frac{d \Omega_j}{4 \pi}
  W_{a b}^j U_{a b j} ( L ) \left[ g_{a j} ( L ) g_{j b}
  ( L ) - g_{a b} ( L ) \right]
\end{equation}
In this form, the second term on the right-hand side only contributes
to the NGLs starting at order $L^3$. If we ignore this term, the BMS equation
reduces to a linear differential equation which is straightforward to solve.
For the opposite-hemisphere case, with $a=n$ and $b=\nbar$,  we get
\begin{align}
  \partial_\tL  g^{(2R)}_{n \nbar} ( L ) &=  g_{n \nbar}^{(2R)} (L)
  \int_{\text{left}}  \frac{d \Omega_j}{4 \pi} W_{n \nbar}^j \left[ U_{n
  \nbar j} ( L ) - 1 \right]
\\
  &= 2^\tL \int_0^1 d c \frac{1}{( 1 - c^2 )} \left[ \frac{2^\tL
  c^\tL}{( 1 + c )^\tL} - 1 \right]  g_{n \nbar}^{(2R)} (L)
\\
  &= - \frac{1}{2} \left( \gamma_E + \frac{\Gamma' ( \tL )}{\Gamma  ( \tL )} + \frac{1}{\tL} \right) 
  g_{n \nbar}^{(2R)} (L)
\end{align}
with $\Gamma ( L )$ the gamma function. The solution to this differential equation is
\begin{equation}
  g^{(2R)}_{n \nbar} = \sqrt{\frac{e^{- \gamma_E \tL}}{\Gamma ( 1 + \tL )}}
  =1-\frac{\pi^2}{24} \tL^2 +\frac{\zeta(3)}{6}  \tL^3+\cdots
  \label{gnnfactor}
\end{equation}

This partially resummed result is not particularly useful, as it does not dominate the full solution in any particular limit. It nevertheless has some interesting features:
\begin{itemize}
\item Unlike the naive exponentiation of the 2-loop results, $g_{ab}  = \exp(-\frac{\pi^2}{24} \tL^2)$,
the resummed 2-loop result includes odd powers of $\tL$ in its expansion.
\item The expansion of this result contains half of the 3-loop leading NGL.
\item There is an intriguing formal similarity between this solution and solutions to renormalization
group equations for global logarithms (see e.g.~\cite{0803.0342}). These RGEs are easiest to solve
in Laplace space. This suggests there may be a way to solve the BMS equation exactly using some clever integral transform. 
\end{itemize}
A comparison of $  g^{(2R)}_{n \nbar}$, $g_{ab}  = \exp(-\frac{\pi^2}{24} \tL^2)$, the numerically resummed result, and the 5-loop approximation are shown  in Fig.~\ref{fig:gnn}.

\newpage

\subsection{Numerical resummation}

\begin{figure}[t]
\begin{center}
  \includegraphics[width=0.45\textwidth]{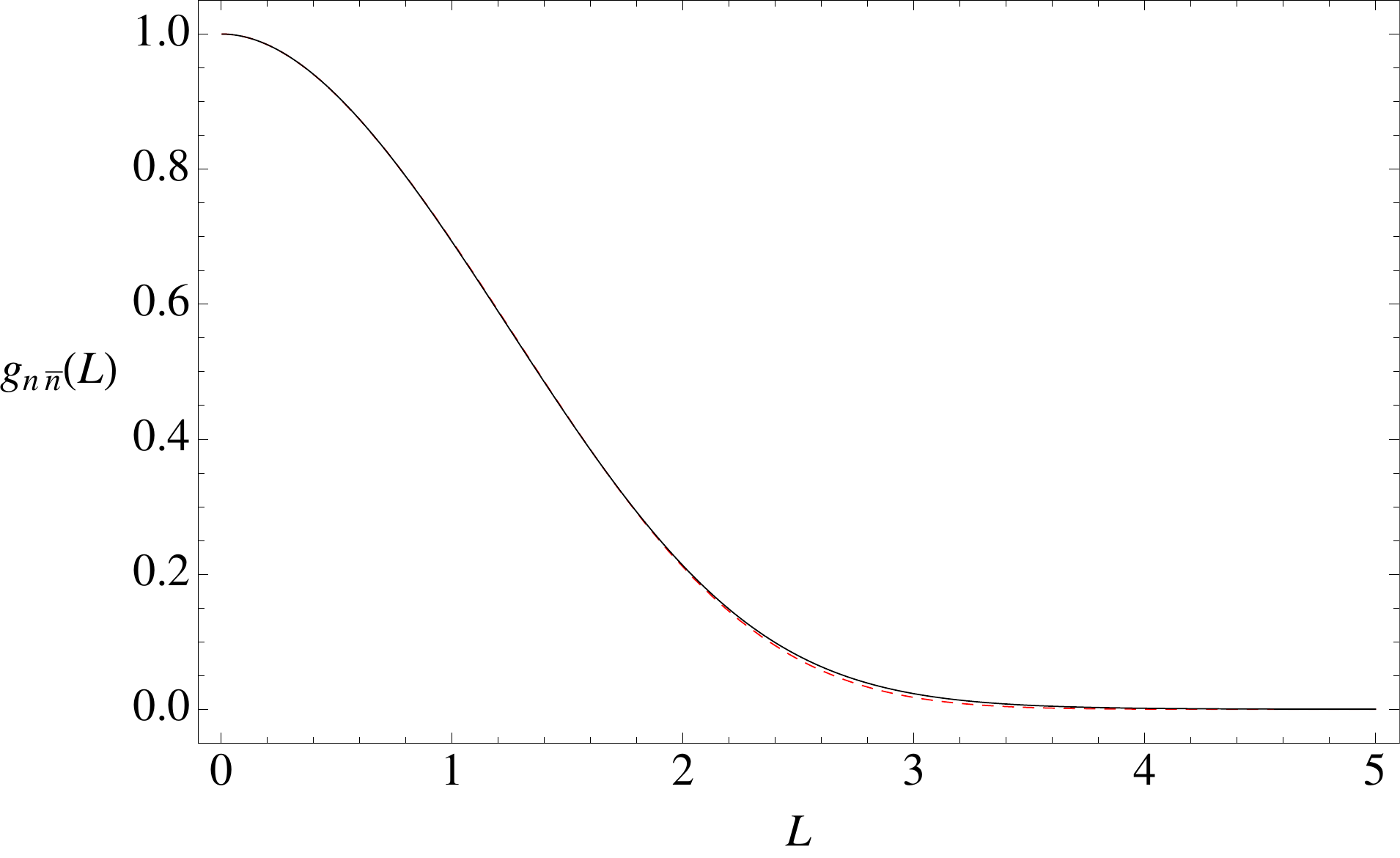}
 \includegraphics[width=0.45\textwidth]{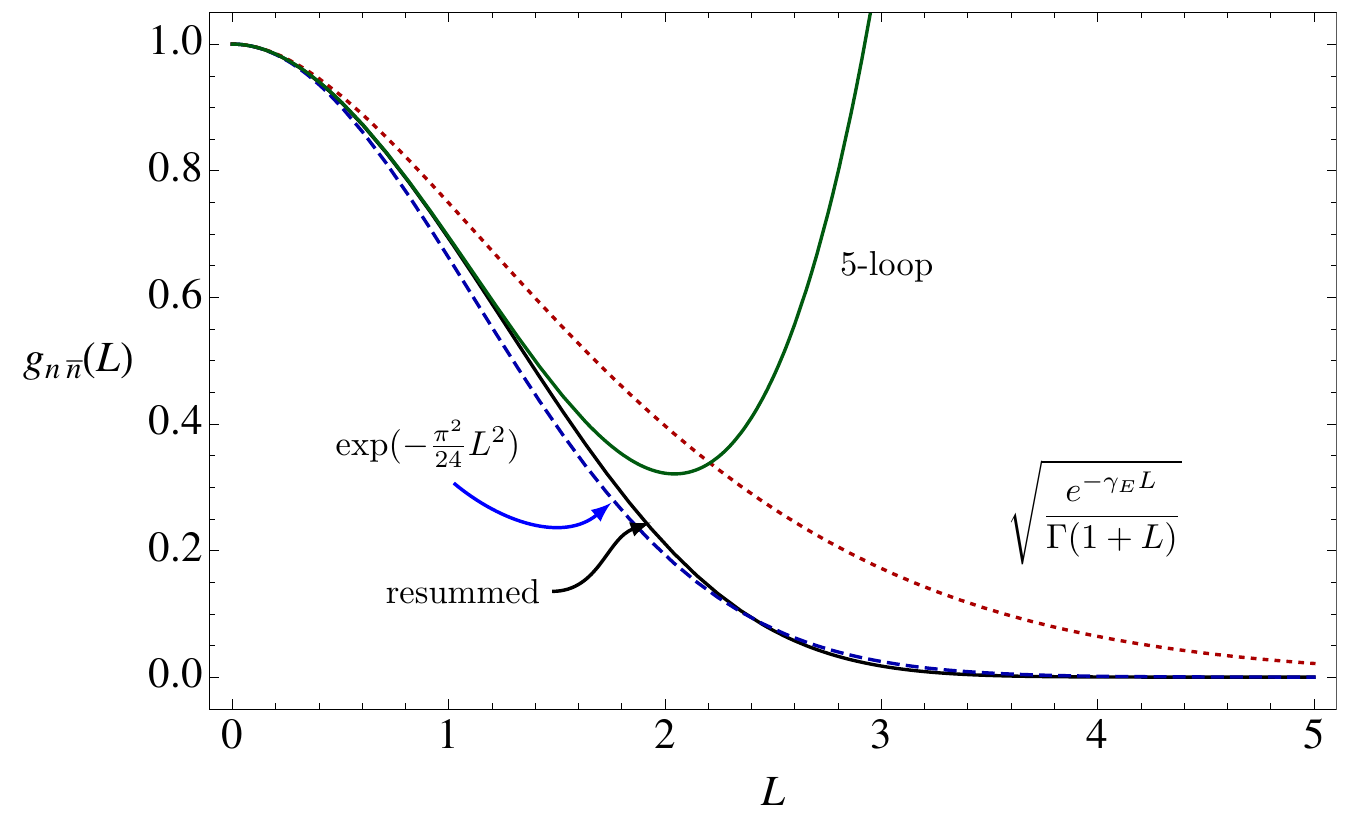}  

  \caption{ Left shows a comparisons of the resummation of the leading
    NGL using the Monte Carlo approach of  Dasgupta and Salam (dashed,
    red) to a numerical integration of the BMS equation (black). Right
    shows a comparison of the numerically integrated result to various
    approximations.
  }
  \label{fig:gnn} 
\end{center}
\end{figure}

The BMS equation for the non-global logarithms, in the form of Eqs.
(\ref{BMSab}) and (\ref{BMSan}) can be solved numerically for $g_{ab}(L)$.
Since this equation has only a single derivative, and the boundary condition $g_{ab}(0)=1$ for all $a,b$ is simple, we can solve the equation by simply integrating. 
As noted in Section~\ref{sec:bmsprop}, although $a$ and $b$ are points on the sphere, 
$g_{ab}$ only depends
on the invariant distance $\met{ab}$ associated with the \poincare disk after stereographic
projection. Rather than exploiting this, we take a more brute-force approach
and use only the obvious azimuthal asymmetry: we parameterize $a$ and $b$ by
  by $\cos \theta_a$,
$\cos \theta_b$ and $\phi_a - \phi_b$.
To solve the BMS equation numerically, we discretize the angles
into $n_{\theta}$ and $n_{\phi}$ bins, and solve the equation by summing the integrand with step size
$\Delta L = \tL/n_L$. The computation time using this approach scales like $ n_L n_c^3 n_\phi^2$.

The only thing which makes the numerical integration nontrivial is the collinear
singularity when $j = a$ of $j = b$. This singularity causes no problem in an analytic integral (it
can be integrated over), but must be avoided in a discretized approach. We
take the simplest solution and  simply omit the $j = a$ and
$j = b$ bins. This omission obviously affects the results for finite $n_c$, but smoothly
disappears as $n_c \to \infty$,  Rather than trying to take very large $n_c$, we simply
take values of order $n_c = 30$ or $n_c = 40$ and extrapolate to $n_c = \infty$ from a fit as a
function of $n_c$.

Our solution for $g_{n \nbar} (L)$  is shown in Fig.~\ref{fig:gnn}. On the left side of this figure, it
is compared to   the numerical calculation of the same quantity by Dasgupta and Salam ~\cite{Dasgupta:2001sh}. More precisely, we compare to the fit given in their paper, which in our normalization is
\begin{equation}
g_{n\nbar}^{\text{(DS)}}(L)=\exp\left[-\frac{\pi ^2}{24} \tL^2 \frac{1+ 0.180625 \tL^2}{1+ 0.325472 \tL^{1.33}}\right]
\end{equation}
 In the region $L<1.4$, where the
fit in~\cite{Dasgupta:2001sh} is claimed to be valid, we find less than a $0.1\%$ disparity.
This confirms the equivalence of the two approaches. It is notable
however that $g_{n\nbar}^{\text{(DS)}}(L)$ does not have a series expansion around $\tL=0$. Indeed,
at $\tL=0$, the third derivative of this function is zero  and the fourth derivative is infinite.

The right side of Fig.~\ref{fig:gnn} compares the resummed distribution to various approximations.
It is very interesting that the exponential of the 2-loop leading NGL provides the best approximation. We have no explanation of this fact.  

Figure~\ref{fig:conv} compares the resummed distribution to the $n$-loop approximation,
with $n=2,3,4$ and $5$. The series appears not be convergent beyond $\tL \approx 1$.  

\section{Non-global logarithms at finite $N_c$}
\label{sec:finitenc}
We have discussed the equivalence of explicit
matrix element construction of NGLs and the iterative expansion of BMS
equation at large $N_c$. We have also discussed the symmetry of
the BMS equation. In this section, we  briefly discuss the implication of combining
these two ingredients at finite $N_c$. 

An evolution equation describing the evolution of NGLs at finite $N_c$
has been conjectured by Weigert~\cite{hep-ph/0312050} and solved
numerically by Hatta and Ueda~\cite{1304.6930}. This conjecture is based on the
formal similarity of the BMS equation and BK equation, and that the BK
equation has a finite $N_c$ generalization, the JIMWLK
equation~\cite{hep-ph/9606337,hep-ph/9706377,hep-ph/0102009}. However,
to the best of our knowledge, there is no direct derivation of this evolution equation, nor
is there a proof that the equation reproduces all of the leading NGLs at finite $N_c$.

It is certainly true that the SEO approximation can be applied at finite $N_c$. 
Instead of dipoles, more involved recursive soft gluon insertion formulas
must be used,  but one still has physical picture of Wilson lines emitting gluons which then become 
new Wilson lines. Thus the matrix element construction of the NGL integrand  described in
Sec.~\ref{sec:seo} and \ref{sec:nghmi} still applies. 
A formulation for the recursive picture is discussed in~\cite{Bassetto:1984ik,hep-ph/9908523}
using
the recursive soft gluon insertion formula in 
color space
notation~\cite{Bassetto:1984ik,hep-ph/9908523}. Specifically, the
squared amplitudes at finite $N_c$ for $m+1$ real soft gluon emissions with SEO
 from a quark-antiquark dipole $(ab)$ can be obtained from the corresponding
amplitudes with the softest gluon removed:
\begin{align}
  \left|M^{1,\dots,m}_{ab}\right|^2 = -g^2 \sum_{i,j=a,b,1,\dots,m-1}
  \int \frac{\omega_{m} d\omega_{m}}{4\pi^2}
  \frac{d\Omega_{m}}{4\pi}
  \frac{(ij)}{(im)(mj)}
    \left|M^{(i,j);1,\dots,m-1}_{ab} \right|^2,
\label{eq:fullc}
\end{align}
where we have also included the soft phase space factor. The color
correlated amplitude $M^{(i,j);1,\dots,m}_{ab}$ is defined as~\cite{hep-ph/9908523}
\begin{align}
 \left| M^{(i,j);1,\dots,m}_{ab}\right|^2  = \langle M^{1,\dots,m}_{ab}
  |\boldsymbol{T}_i \cdot \boldsymbol{T}_j| M^{1,\dots,m}_{ab} \rangle
\end{align}
To make contact with the leading color-squared amplitudes,
Eq.~(\ref{generalM}), we note that the color correlation factorizes
into individual color dipoles at large $N_c$. In particular, for a
color dipole $(ij)$, we have
\begin{align}
  \boldsymbol{T}_i + \boldsymbol{T}_j \overset{N_c \to \infty}{=} 0 \Rightarrow \boldsymbol{T}_i
  \cdot \boldsymbol{T}_j = 
\left\{ \begin{array}{l l}
     -\frac{N_c}{2} & \quad i,j=a,b
\\
   -N_c & \quad \text{otherwise}
  \end{array}
\right. 
\end{align}
where the factor of $1/2$ in the first case comes from the fact that
$C_F = \frac{N_c}{2} + \mathcal{O}(1/N_c)$.

Although the color structure is more complicated at finite $N_c$, the kinematics of factorization formula in Eq.~\eqref{eq:fullc} is not. In particular, the energy integral still trivially factorizes off and
the radiator function $\cW_{ij}^m =\frac{ (ij) }{ (im ) (mj) } $ is unchanged.
Thus, the real emission integrand still enjoys the full  M\"obius symmetry, $\pslc$ of the BMS equation.
 Following the discussion in Sec.~\ref{sec:seo},
it is also easy to see that, since the real-virtual and virtual
corrections in the SEO limit are of the same form as the real emission, they
also preserve the symmetry. 
As at finite $N_c$,  the integration region for the hemisphere NGLs, 
breaks $\pslc$ down to $\pslr$ symmetry of the \poincare disk.
There is no clear reason why this $\pslr$ symmetry should be further
broken. Thus we expect that even at finite $N_c$, $g_{ab}(L)$ depends
on $a$ and $b$ only though the invariant $\met{ab}$.

\section{Conclusions}
\label{sec:conclude}
In this paper we have explored the structure of the leading non-global logarithmic series for
the  hemisphere mass distribution in $e^+e^-$ collisions. The NGLs are represented by
functions $g_{ab}(L)$, where $a$ and $b$ are directions of hard colored particles
producing the radiation which generates the NGLs. For the $e^+e^-$ case
we take $a$ and $b$ aligned with the hemisphere axis $n$ and $\nbar$, and only
 $g_{n \nbar}(L)$ is relevant.
In the general case, $a$ and $b$ do not have to be aligned with the hemisphere axis $n$. We
consider the more general $g_{ab}(L)$ since it feeds in to $g_{ n\nbar}$ and since it has
interesting symmetry properties.

The leading NGLs can be computed using the strong-energy ordering (SEO) approximation. This approximation simplifies both the real-emission matrix elements as well as contributions to the cross section involving virtual and real-virtual graphs. The SEO has led to the BMS equation~\cite{Banfi:2002hw} which allows for the resummation of the leading NGL. We checked that the real/virtual contributions to the cross section for NGLs agree with the expansion of the BMS equation order-by-order in perturbation theory. 

One advantage of using the BMS equation is that it manifests most clearly the $\pslc$ symmetry observed in~\cite{0903.4285,0909.0056}. We showed that the hemisphere integration region breaks the $\pslc$ symmetry down to $\pslr$ which is the isometry group of the \poincare disk. Angles on the hemisphere representing ends of a color dipole furnish a representation of $\pslr$ which can be constructed through a stereographic projection onto the equatorial disk. The result is that $g_{ab}(L)$
only depends on a single invariant, the distance $\met{ab} = \frac{1-\cos \theta_{ab} } {\cos \theta_a \cos\theta_b}$ between $a$ and $b$, where $\theta_a$ and $\theta_b$ are the polar angles with respect the hemisphere axis and $\theta_{ab}$ is the angle between them.
This invariant greatly simplifies the calculation of the leading NGLs at fixed order.

To compute the fixed-order expansion of the leading hemisphere NGLs,
we iterated the BMS equation. At each loop order, only one azimuthal
angle and one polar angle integral needs to be done. The azimuthal
integrals are straightforward to do by deforming the integration
contour. The result at $n$ loops is a set
of classical polylogarithms of uniform transcendentality weight
$n$. To do the polar angle integrals, we convert these classical
polylogarithms into Goncharov polylogarithms in a canonical form using
the tensor algebra of the symbol. The symbols gives the complete result
up to constants of uniform transcendentality, like $\zeta(n)$ or
$\pi^n$. At 3 loops, these constants can be guessed, but at 4 loops we
require the use of the coproduct to extract them. The result is a
formula for $g_{ab}(L)$ at 4 loops. We then use this formula to
compute $g_{n\nbar}(L)$ at 5 loops. This is our main concrete result, given in Eq.~\eqref{fiveloop}.

In addition to computing the leading NGL at 5 loops, we resummed the leading NGL to all orders by solving the BMS equation numerically. We found a result in very good agreement with the fit from a Monte Carlo calculation presented in~\cite{Dasgupta:2001sh}, confirming the equivalence of the
BMS equation and the Monte Carlo approach. Interestingly, the resummed distribution seems
to agree quite well with the exponentiation of the 2-loop result, despite the apparent importance of the 3-loop NGL coefficient.

The 5-loop leading hemisphere NGL may be of some (limited) phenomenological importance, since
it contributes to event shapes like the heavy jet mass~\cite{1005.1644}. However, more profound consequences of this work probably include the relatively simple structure of the leading NGL series. Working in the strong-energy-ordered approximation apparently produces an extended symmetry  which is only partially broken through a finite integration region. That the NGLs are computed with iterated integrals of uniform transcendentality is also somewhat surprising. While such integral series are common in supersymmetric settings, examples in large $N_c$ QCD ($\mathcal{N}=0$) are more rare. It may be important to understand the symmetry and the generality of the iterated structure in more depth.

\section*{Acknowledgments}
The authors would like to thank A. Banfi, L. Dixon, M.
Dasgupta, I. Feige, E. Gardi, G. Salam and R. Schabinger for useful
discussions. HXZ also thanks L. Dixon for enlightening discussion
on the perturbation convergence of Mueller-Navelet jet~\cite{1309.6647}, and to A. von Manteuffel for sharing his
private code for numeric evaluation of GPLs using \texttt{GiNaC}~\cite{cs/0004015}.
MDS is supported by the Department of Energy, under grant
DE-SC003916. HXZ is supported by the Department of Energy under
contract DEAC0276SF00515.

\appendix
\section{Azimuthal integrals}
\label{app:azi}

We present some useful formulae for azimuthal integral in this
appendix. The 1 and 2-loop results are simple
\begin{equation}
\az_1=  \int^{2\pi}_0 \frac{d\phi_j}{2\pi} \frac{1}{(jb)} =
  \frac{1}{|\cos\theta_j - \cos\theta_b|},
\end{equation}  
\begin{equation}
\az_2=  \int^{2\pi}_0 \frac{d\phi_j}{2\pi} \frac{\ln(1+\met{jb})}{(jb)} = 
  \frac{1}{ \cos\theta_j - \cos\theta_b} \ln
  \frac{1+\met{b\nbar}}{1+\met{j\nbar}},
\end{equation}  
The 3-loop result is more complicated
\begin{multline}
\az_3=  \int^{2\pi}_0 \frac{d\phi_j}{2\pi} \frac{\ln \met{jb} \ln(1+\met{jb}) +
    \mathrm{Li}_2 (-\met{jb})}{(jb)}  = \frac{1}{\cos\theta_b -
    \cos\theta_j} \\
    \times \left[ \ln\frac{1+\met{j\nbar}}{1+\met{b\nbar}} \ln
    \frac{ \met{j\nbar}+ \met{b\nbar} + |\met{j\nbar} - \met{b\nbar}
      |}{2} + \mathrm{Li}_2(-\met{j\nbar}) - \mathrm{Li}_2 (-\met{b\nbar})
\right].\label{eq:azi3}
\end{multline}
At 4 loops the result is most usefully expressed in terms of GPLs in canonical form
\begin{align}
\az_4= & (\cos\theta_b - \cos\theta_j)  \int^{2\pi}_0 \frac{d\phi_j}{2\pi (jb)}  \left[ - \frac{1}{6}
    r^3_{jb} + \frac{1}{L^3} g^{(3)}_{jb}(L) - \frac{1}{L^2} r_{jb}
    g^{(2)}_{jb}(L) \right]
\nn
\\
=& 
- \frac{1}{12} G(-1,0,\met{b\nbar}; \met{j\nbar}) - \frac{1}{12}
G(0,-1,\met{b\nbar}; \met{j\nbar}) + \frac{1}{8} G(-1,0,-1;
\met{j\nbar})
\nn
\\
&
- \frac{1}{24} G(0,-1; \met{j\nbar}) G(-1; \met{b\nbar}) 
- \frac{1}{12} G(0,-1;\met{j\nbar}) G(0; \met{b\nbar}) + \frac{1}{12}
G(-1; \met{j\nbar}) G(0,0; \met{b\nbar})
\nn
\\
&
+ \frac{1}{24} G(-1; \met{j\nbar})G(0,-1;\met{b\nbar}) - \frac{1}{12}
G(-1,0,0; \met{b\nbar}) - \frac{1}{24} G(-1,0,-1; \met{b\nbar}) 
\nn
\\
&
+ \frac{\pi^2}{36} G(-1; \met{b\nbar}) - \frac{\pi^2}{36} G(-1;\met{j\nbar}),
\label{eq:azi4}
\end{align}
which is valid for $\met{j\nbar} > \met{b\nbar}$. This equation is in the canonical GPL form,
since the next integration variable $\met{j\nbar}$  shows up only in
the argument of GPLs. Some details of how this canonical form is realized are explained in Appendix~\ref{app:coproduct}.

\section{General hemisphere NGL functions to 4 loops}
\label{app:gab}
 For
the same hemisphere NGLs, that is, both $a$ and $b$ are in the left
hemisphere, we have obtained the analytical results up to and include
four loops. Defining $x = \met{ab}$ we find
\begin{align}
  \frac{1}{\tL^2}g^{(2)}_{ab}(L) =& - \frac{1}{4} G(-1,-1;x) +
  \frac{1}{4} G(-1,0;x)
\nn
\\
=& \frac{1}{4} \ln x \ln(1+x) - \frac{1}{8} \ln^2 (1+x) +
\mathrm{Li}_2(-x),
\\
\frac{1}{\tL^3} g^{(3)}_{ab}(L) = & \frac{\pi^2}{36} G(-1;x) -
\frac{1}{4} G(-1,-1,-1;x) + \frac{1}{4} G(-1,-1,0;x) + \frac{1}{12}
G(-1,0,-1;x)
\nn
\\
& - \frac{1}{12}G(-1,0,0;x)
\nn
\\
=& \frac{\pi^2}{72} \ln(1+x) - \frac{1}{24} \ln^2 x\ln(1+x) +
\frac{1}{12} \ln x \ln^2 (1+x) - \frac{1}{36} \ln^3 (1+x)
\nn
\\
&
- \frac{1}{12} \ln x \mathrm{Li}_2(-x) + \frac{1}{12} \ln(1+x)
\mathrm{Li}_2 (-x) + \frac{1}{12} \mathrm{Li}_3(-x) - \frac{1}{12}
\mathrm{Li}_3 \left(\frac{1}{1+x} \right) + \frac{\zeta(3)}{12},
\\[10mm]
\frac{1}{\tL^4} g^{(4)}_{ab}(L) = & \frac{\pi^2}{36} G(-1,-1;x) -
\frac{\pi^2}{144} G(-1,0;x) - \frac{3}{16} G(-1,-1,-1,-1;x) +
\frac{3}{16} G(-1,-1,-1,0;x)
\nn
\\
& + \frac{1}{12} G(-1,-1,0,-1;x) -
\frac{1}{12} G(-1,-1,0,0;x) + \frac{1}{48} G(-1,0,-1,-1;x)
\nn
\\
&
- \frac{1}{96} G(-1,0,-1,0;x) - \frac{1}{32} G(-1,0,0,-1;x) +
\frac{1}{48} G(-1,0,0,0;x)
- \frac{\zeta(3)}{16} G(-1;x)
\nn
\\
=& -\frac{11\pi^2}{576} \ln x\ln(1+x) + \frac{1}{288} \ln^3 x
\ln(1+x) + \frac{\pi^2}{72} \ln^2 (1+x)
\nn
\\
&
 + \frac{1}{24} \ln(-x)\ln(x)
\ln^2(1+x) - \frac{1}{48} \ln^2 x \ln^2(1+x) + \frac{1}{48} \ln(-x)
\ln^3(1+x) 
\nn
\\
&
+ \frac{1}{32} \ln x \ln^3(1+x) - \frac{1}{128} \ln^4(1+x) -
\frac{\pi^2}{144} \mathrm{Li}_2(-x) + \frac{1}{96} \ln^2 x
\mathrm{Li}_2 (-x) 
\nn
\\
&
+ \frac{5}{96} \ln^2(1+x) \mathrm{Li}_2 (-x) + \frac{1}{96}
\mathrm{Li}_2 (-x)^2 + \frac{1}{96} \ln x \ln(1+x) \mathrm{Li}_2(1+x)
\nn
\\
&
+ \frac{1}{24} \ln^2(1+x) \mathrm{Li}_2(1+x) - \frac{1}{48} \ln x
\mathrm{Li}_3 (-x) + \frac{3}{32} \ln(1+x) \mathrm{Li}_3(-x) 
\nn
\\
&
+ \frac{1}{16} \ln x \mathrm{Li}_3(1+x) - \frac{1}{24} \ln(1+x)
\mathrm{Li}_3 (1+x) + \frac{1}{48} \mathrm{Li}_4(-x) + \frac{1}{16}
S_{2,2}(-x) 
\nn
\\
&
- \frac{\zeta(3)}{16} \ln x - \frac{\zeta(3)}{48} \ln(1+x),
\end{align}
We have given separately the GPL
representation and classical polylogarithms representation
of the results. The classical polylogarithms representation is
obtained using the package \texttt{HPL}~\cite{Maitre:2005uu}. 

The hemisphere NGL functions $g_{a n}(L)$ when one direction is in the right-hemisphere are given in Eq.~\eqref{nglLR2} to \eqref{nglLR4} .

\section{Systematic use of the symbols and coproducts}
\label{app:coproduct}
In this Section we explain how the form of Eq.~(\ref{eq:azi4}) which is canonical in terms
of GPLs is obtained. The azimuthal integral
\begin{equation}
\az_4=  (\cos\theta_b - \cos\theta_j)  \int^{2\pi}_0 \frac{d\phi_j}{2\pi (jb)}  \left[ - \frac{1}{6}
    r^3_{jb} + \frac{1}{L^3} g^{(3)}_{jb}(L) - \frac{1}{L^2} r_{jb}
    g^{(2)}_{jb}(L) \right]
\end{equation}
 can be done using the contour integral method sketched in Section~\ref{sec:ngl2}
 and  \texttt{Mathematica}. However, the resulting expression is very complicated and
further integrating over $\met{j\nbar}$ is too difficult.
However, the symbol of $\Phi_4$ is not unmanageable:
\newpage
\begin{align}
  \mathcal{S}[\Phi_4] = & \frac{1}{12} \met{b\nbar} \otimes
  (1+\met{b\nbar}) \otimes \met{b\nbar} - \frac{1}{12} \met{b\nbar}
  \otimes (1+\met{b\nbar}) \otimes \met{j\nbar} - \frac{1}{12}
  \met{b\nbar} \otimes \met{j\nbar} \otimes (1+\met{b\nbar})
\nn
\\
&
+
\frac{1}{12} \met{b\nbar} \otimes \met{j\nbar} \otimes
(1+\met{j\nbar})
- \frac{1}{24} (1+\met{b\nbar}) \otimes \met{b\nbar} \otimes
(1+\met{b\nbar})
\nn
\\
&
 + \frac{1}{24} (1+\met{b\nbar}) \otimes \met{b\nbar}
\otimes (1+\met{j\nbar})
+ \frac{1}{24} (1+\met{b\nbar}) \otimes (1 + \met{j\nbar}) \otimes
\met{b\nbar}
\nn
\\
&
- \frac{1}{24} (1+\met{b\nbar}) \otimes (1+\met{j\nbar}) \otimes
\met{j\nbar}
- \frac{1}{12} (\met{b\nbar} - \met{j\nbar}) \otimes \met{b\nbar}
\otimes (1+\met{b\nbar})
\nn
\\
&
+ \frac{1}{12} (\met{b\nbar} - \met{j\nbar}) \otimes \met{b\nbar}
\otimes (1+ \met{j\nbar})
-
\frac{1}{12} (\met{b\nbar}-\met{j\nbar}) \otimes (1+\met{j\nbar})
\otimes \met{b\nbar}
\nn
\\
&
+ \frac{1}{12} (\met{b\nbar} - \met{j\nbar}) \otimes (1+\met{b\nbar})
\otimes \met{j\nbar}
+ \frac{1}{12} (\met{b\nbar} - \met{j\nbar}) \otimes \met{j\nbar}
\otimes (1+\met{b\nbar})
\nn
\\
&
- \frac{1}{12} (\met{b\nbar} - \met{j\nbar}) \otimes \met{j\nbar}
\otimes (1+ \met{j\nbar})
+ \frac{1}{12} (\met{b\nbar} - \met{j\nbar}) \otimes (1+ \met{j\nbar})
\otimes \met{b\nbar} 
\nn
\\
&
- \frac{1}{12} (\met{b\nbar} - \met{j\nbar}) \otimes (1+ \met{j\nbar})
\otimes \met{j\nbar} - \frac{1}{8}  (1+\met{j\nbar}) \otimes
(1+\met{b\nbar}) \otimes \met{b\nbar} 
\nn
\\
& - \frac{1}{8} (1+\met{j\nbar})
\otimes (1+ \met{b\nbar}) \otimes \met{j\nbar}
- \frac{1}{8} (1+\met{j\nbar}) \otimes \met{j\nbar} \otimes
(1+\met{b\nbar})
\nn
\\
&
+ \frac{1}{8} (1+\met{j\nbar}) \otimes \met{j\nbar} \otimes (1+\met{j\nbar}).
\end{align}

From the symbol, we can
reconstruct the  {\it  most complicated} part of the original
expression, using an algorithm described in Ref.~\cite{1302.4379}. We start with the
terms with the most number of $\met{j\nbar}$ factors. That is, where  $\met{j\nbar}$
 shows up in all the slots, {\it e.g.}, $
(1+\met{j\nbar}) \otimes \met{j\nbar} \otimes
(1+\met{j\nbar})$. For each such term, a GPL that has the same symbol can be immediately
read off from its entries. For example, 
\begin{align}
  \mathcal{S}\Big[G(-1,0,-1; \met{j\nbar})\Big] =(1+\met{j\nbar}) \otimes \met{j\nbar} \otimes
(1+\met{j\nbar}).
\end{align}
In this way, we construct an ansatz $\az^{G1}_4$, consisting of GPLs in the canonical form,
whose symbol exactly matches the  terms in $\mathcal{S}[ \az_4] $ with the most
 $\met{j\nbar}$ factors. 
The symbol of the remainder, $\mathcal{S}( \az_4 - \az^{G1}_4 )$ now contains terms where
at least one of the slot is free of $\met{j\nbar}$, {\it e.g.}, $ (1+\met{j\nbar}) \otimes \met{j\nbar} \otimes
(1+\met{b\nbar})$. An ansatz for some terms in of this form might have the symbol
\begin{multline}
  \mathcal{S}\Big[ G(0,-1; \met{j\nbar}) G(-1; \met{b\nbar})\Big] = (1 +
  \met{j\nbar} ) \otimes \met{j\nbar} \otimes (1+\met{b\nbar}) \\
  +
  (1+\met{j\nbar}) \otimes (1+\met{b\nbar}) \otimes \met{j\nbar} 
+ (1+\met{b\nbar}) \otimes (1+\met{j\nbar}) \otimes \met{j\nbar}.
\end{multline}
Organizing the matching in this way systematically leads to a guess $\az_4^{G}$ with
GPLs in canonical form with the same symbols as $\az_4$.

Since $\az_4$ and $\az_4^G$ have the same symbol, they can only differ
 by constants of the appropriate transcendentality.  For transcendentality-weight $3$ GPLs, the terms missed from the symbol construction can only either be $\zeta(3)$ or could be $\zeta(2)\times 
 \ln[ R(\met{bn}) ]$ for rational functions $R(x)$. 
 The terms
proportional to $\zeta(2)$ can be extracted by the coproducts
$\Delta_{2,1}$ \cite{1203.0454}. Specifically, we have
\begin{align}
  \Delta_{2,1} \Big[ \az_4 - \az^G_4 \Big]= \frac{1}{6}\Big( \zeta(2)\otimes
  \ln(1+\met{b\nbar})\Big) - \frac{1}{6}\Big( \zeta(2) \otimes \ln (1+ \met{j\nbar})\Big).
\end{align}
The action of $\Delta_{2,1}$  fixes the terms proportional to $\zeta(2)$. It suggests
that we should add
\begin{equation}
\Delta \az_4^G =  \frac{\zeta(2)}{6} G(-1; \met{b\nbar}) -\frac{\zeta(2)}{6}  G(-1;\met{j\nbar}),
\end{equation}
to our guess. 
 Finally, one can check at a random phase space
point that the difference of $\az_4$ and $\az^G_4 + \Delta \az_4^G$ vanishes,
showing that there is no missing $\zeta(3)$ term. The result is  Eq.~\eqref{eq:azi4}.

\bibliographystyle{JHEP3}

\end{document}